\documentclass[printer]{aa}
\usepackage[utf8]{inputenc}
\usepackage{epsfig}
\usepackage{amsmath}
\usepackage{amsfonts}
\usepackage{amssymb}
\usepackage{mathtools}
\usepackage{graphicx}
\usepackage{colordvi}
\usepackage{color}
\usepackage{acronym}
\usepackage[normalem]{ulem}
\usepackage[varg]{txfonts}

\usepackage{hyperref}
\hypersetup{
	colorlinks=true,        
	linkcolor=blue,         
	citecolor=cyan,         
}

\usepackage{upgreek}
\usepackage{natbib}

\newacro{EHT}{Event Horizon Telescope}
\newacro{ngEHT}{new generation Event Horizon Telescope}
\newacro{BH}{black hole}
\newacro{BHT}{black hole -- torus}
\newacro{GRHD}{general relativistic hydrodynamics}
\newacro{GRMHD}{general relativistic magnetohydrodynamic}
\newacro{AGN}{active galactic nucleus}
\acrodefplural{AGN}{active galactic nuclei}
\newacro{ADAF}{advection dominated accretion flow}
\newacro{SANE}{Standard and Normal Evolution}
\newacro{MAD}{Magnetically Arrested Disk}
\newacro{SMBH}{supermassive black hole}
\newacro{MHD}{magnetohydrodynamics}
\newacro{PSD}{power spectral density}
\newacro{GRF}{Gaussian random field}
\newacro{AMR}{Adaptive Mesh Refinement}

\begin{document}
\title{General relativistic hydrodynamic simulations of perturbed transonic accretion}
\titlerunning{Perturbed Transonic Accretion}
\author{H{\'e}ctor R. Olivares S.\inst{1}
	\and Monika A. Mo{\'s}cibrodzka \inst{1}
	\and Oliver Porth \inst{2}}

\institute{Department of Astrophysics/IMAPP, Radboud University Nijmegen, P.O. Box 9010, 6500 GL Nijmegen, The Netherlands \
\and Anton Pannekoek Institute for Astronomy, University of Amsterdam, Science Park 904, 1098 XH Amsterdam, The Netherlands\\
\email{holivares@science.ru.nl}}

\date{
\today
}
\abstract
{Comparison of horizon-scale observations of Sgr~A* and M87*
	with numerical simulations has provided considerable insight
	in their interpretation.
	Most of these simulations are variations of the same physical
	scenario consisting of a rotation
	supported torus seeded with a poloidal magnetic fields.
	However, this approach has several well known limitations such as secular
	decreasing trends in mass accretion rate
	that render long term variability studies difficult,
	a lack of connection with the large-scale accretion flow which
	is replaced by an artificial medium	emulating vacuum, 
	and important differences with respect to the predictions of models of 
	accretion onto Sgr~A* fed by stellar winds.}
{We aim to study the flow patterns that arise at horizon scales in more
	general accretion scenarios, that have a clearer connection with the large
	scale flow and are at the same time controlled by a reduced set of parameters.}
{As a first step in this direction, we
	perform three dimensional general relativistic hydrodynamic simulations
	of rotating transonic flows with velocity perturbations injected from a spherical boundary located far away from the central object (1000 gravitational radii). We study the general properties of these flows with varying angular momentum and perturbation amplitudes.  We analyze time series of mass and angular momentum
	radial fluxes, angle- and time-averaged profiles, and synthetic Bremsstrahlung lightcurves,
	as well as the three-dimensional structure of the flow, and quantify shock- and sonic transitions in the solutions.}
{We observe a rich phenomenology in accretion patterns, that includes smooth Bondi-like flows,
	turbulent torus-like structures, shocks, filaments, and complex sonic structures.
	For sufficiently large perturbations and angular momentum, radial profiles deviate
	from the constant entropy and constant angular momentum profiles
	used for initialization and resemble those of advection dominated accretion flows,
	showing evidence of entropy generation and angular momentum redistribution not mediated by
	magnetic fields.
	Time series do not show the secular decreasing trend and are suitable for long-term variability
	studies. We see that the fluctuations are amplified and extend further in frequency than the injected spectrum, producing a red noise spectrum both for mass accretion rate and
	the synthetic light curves.}
{We present a simulation setup that can produce a wide variety of flow patterns at horizon scales
	and incorporate information from large scale accretion models.
	Future inclusion of magnetic fields and radiative cooling could make this type of
	simulations a viable alternative for numerical modeling of general low-luminosity
	active galactic nuclei.}

\maketitle

\section{Introduction}
\label{intro}

Accretion onto compact objects such as black holes and neutron stars powers some of the most spectacular phenomena in astrophysics.  While the focus of numerous studies in accretion theory is on how matter and angular momentum are transported through an accretion disk, much less studies put into focus the formation of the accretion disk itself.  
In particular numerical simulations of accretion disk formation are encumbered by the large scale separation between circularization radius of incoming matter and the size of the accretor.  
There are certain applications however where the accreted matter has comparatively low angular momentum, leading to circularization radii not much larger than the accretor itself.  Prime examples are the high-mass X-ray binaries (HMXBs) and chaotic stellar wind-fed accretion in galactic nuclei such as our own Galactic center Sgr~A*.   

A central role in the interpretation of the event-horizon-scale observations of Sgr~A* and M87* by the EHT Collaboration is played by general \ac{GRMHD} simulations \citep{PorthChatterjeeEtAl2019a,CollaborationAkiyamaEtAl2019d,CollaborationAkiyamaEtAl2022}.
To date, all of the models in the simulation library for Sgr~A* and most of those used for M87* follow variations of the same initial conditions of a rotation-supported torus \citep{FishboneMoncrief1976} seeded with a weak poloidal magnetic field.  
While a lot of physical insights have already been gained by comparing observational data against GRMHD simulations -- leading to increasingly tight constraints of the parameters such as black hole mass, accretion rate, inclination and black hole spin \citep{CollaborationAkiyamaEtAl2019d,EHT2021,CollaborationAkiyamaEtAl2022} -- there are several limitations intrinsic to the considered ``\ac{BHT}'' simulations.  For example, since they are initialized with a finite amount of matter contained in the torus, the matter content in the simulation decreases over time, accompanied by a corresponding decrease in mass accretion rate.  This secular trend renders the study of long-term variability difficult. This systematic is particularly important since the current set of GRMHD simulations produces highly varying lightcurves which are tightly constrained by the less variable data for Sgr~A* \citep{CollaborationAkiyamaEtAl2022,wielgus_orbital_2022}.

The most important limitation of the \ac{BHT} simulations however concerns physical realism.
It is now widely believed that our galactic center black hole, Sgr~A*, can be fed from the winds of $\sim 30$ massive stars that orbit on the parsec scale \citep{Quataert2004,CuadraNayakshin2008,Ressler2018}.  Whether an accretion disk (torus) forms in this scenario depends not only on the initial wind parameters \citep{Moscibrodzka2006,Shcherbakov2010} but also on the interactions of the unbound winds which can give rise to shocks and hydrodynamic turbulence.  
The flow patterns of realistic stellar wind accretion models for low luminosity \acp{AGN}
differ significantly from the BHT scenario described above. In stellar wind accretion, material forms clumpy structures and has a broad distribution of angular momentum without sufficient time to circularize, and is not generally rotation supported. Instead, it is accreted mainly due to an originally low angular momentum and remains in large part unbound \citep{Ressler2018}.
This latter property is shared by different models of accretion from large scales such as the constant-entropy solutions by
\citet{bondi_spherically_1952,Michel1972,Chakrabarti1996}
and models that include dissipation such as the well-known \acp{ADAF} \citep{Narayan1994}.
Recent \ac{MHD} simulations that focus on the large scale dynamics have revealed further differences to the standard BHT scenario: while magnetic fields from stellar winds are initially weak and passively advected, at horizon scales they accumulate and become dynamically important and start to regulate accretion in a way similar to \acp{MAD} \citep{ResslerQuataertEtAl2020,ressler_ab_2020}.  The lack of a predominant angular momentum or magnetic field direction leads to erratic changes in the orientation
of the accretion disk \citep{ResslerQuataertEtAl2021}. Similar transient behavior can be seen in the direction and power of the jet before the formation of a steady jet \citep{LalakosGottliebEtAl2022}. Simulations of accretion from kpc scales onto the \ac{BH} event horizon have also shown that the
accretion flow in elliptic galaxies as M87 can acquire a variety of patterns that range from rotation-supported disks to chaotic streams \citep{guo_toward_2022}.

In general, simulation-based studies of the horizon-scale structure of the accretion flow resulting from large-scale feeding
present the computational challenge of having to simulate length and time scales spanning $\sim 6$ orders of magnitude,
or dealing with uncertain factors such as the details of stellar winds astrophysics.
It would be therefore desirable to gain more insight on the properties of the accretion flow from the study
of transonic solutions connecting the event horizon to infinity, in a similar manner as the theory of accretion disks
has benefited from the study of analytic solutions for fluids in circular motion around black holes.
Depending on the specific angular momentum and energy, analytic studies of trans-sonic low angular momentum accretion flows \citep[e.g.][]{Fukue1987,Chakrabarti1989,Chakrabarti1996,ChakrabartiDas2004} have revealed different regimes characterized by smooth Bondi-like flows, standing accretion shocks or the formation of circularized tori.  Furthermore, the solutions have been studied including the effects of viscosity \citep{ChakrabartiMolteni1995,LanzafameMolteniEtAl1998}, radiative cooling \citep{MolteniSponholzEtAl1996,OkudaTeresiEtAl2004} and magnetic fields \citep{Proga2003b,OkudaSinghEtAl2019,MitraMaityEtAl2022a}, often with particular focus on the stability and dynamics of the accretion shock.  
Numerical simulations of transonic hydrodynamic solutions were presented more recently by \cite{KimGarainEtAl2017,KimGarainEtAl2019} for the Schwarzschild and Kerr spacetimes,
showing that certain perturbations can trigger long-surviving shocks at the location of predicted standing shocks \citep{Chakrabarti1996}.

In fact, a realistic approach to the problem should consider the destabilizing effect of inhomogeneities in the surrounding medium.
The stability of spherical Bondi accretion has been studied analytically in a number of works; see for instance \citet{Moncrief1980} and \citet{kovalenko_instability_1998}.
In the latter work, it is shown that this solution is unstable for non-radial perturbations,
although for the instability to manifest itself the size of the accretor needs to be sufficiently small compared to the Bondi radius,
precisely as it is the case for the nearest \acp{SMBH}.

In this paper, we study a simulation setup
that aims to address the above described limitations of the \ac{BHT} paradigm and
to facilitate the incorporation of information gained from larger scale simulations.
This model depends on a reduced set of parameters that can in principle be chosen to match the properties
inferred for known \acp{SMBH} such as Sgr A*, M87, and other targets of the \ac{EHT} and the planned \ac{ngEHT}.
By incorporating time-dependent properties of the surrounding medium in the boundary conditions,
the simulation domain can be of a modest size comparable to that of existing \ac{GRMHD} simulations
in the \ac{EHT} library.
The simulations presented here are run in pure \ac{GRHD}, that is, with zero magnetic field,
as an intermediate step towards \ac{GRMHD} simulations that will be presented in a forthcoming work.
We show that the proposed setup produces steady time series that are in principle suitable for
long-term variability studies, and exhibits a rich phenomenology that can differ significantly
both from typical \ac{BHT} simulations and from unperturbed Bondi-like accretion.

We describe this setup in Section \ref{sec:setup} and provide a justification for the physical
parameters employed (Section \ref{sec:phys-params}).
In Section \ref{sec:results}, we report on the properties observed in the simulations,
such as time series of mass and angular momentum accretion rates and radial profiles
(Section \ref{sec:global-properties}), three-dimensional morphology, including the presence
of shocks and complex sonic structures (Section \ref{sec:3D-morphology}) and
variability properties \ref{sec:observable}. We summarize and discuss our results
in Section \ref{sec:discussion}, and complement this work with more information
on the simulation setup in the Appendices.


\section{Simulation setup}
\label{sec:setup}
To explore the flow patterns arising from transonic accretion
of an inhomogeneous interstellar medium, we perform three-dimensional \ac{GRHD}
simulations that continuously inject matter from an outer boundary.  
We employ units such that $G=c=1$, so that the gravitational radius
$r_g=GM/c^2$ and the gravitational timescale $t_g=r_g/c$ are $r_g=t_g=M$,
where $M$ is the mass of the black hole.
We adopt a Kerr spacetime with dimensionless spin parameter $a\coloneqq J/M = 0.95$
and the event horizon located at $r_{\rm H}/M = 1 + \sqrt{1 - a^2}$.
For all of our simulations, we set the sonic radius to $r_s=500\ M$
and place the boundary at $r=1000\ M$.
We initialize the simulations with a smooth quasi-stationary background solution with a
latitude-dependent angular momentum profile.  Following \cite{Proga2003b} we adopt an angular momentum profile
that peaks at the equator and vanishes at the poles
(see Appendix~\ref{sec:background} 
for a detailed discussion of the background flow):

\begin{equation}
	\label{eq:ell-profile}
	\ell (\theta) = \ell_0 (1 - |\cos \theta |).
\end{equation}
The background flow is characterized by only two parameters, the angular momentum at
the equator $\ell_0$, and the sonic radius (or alternatively fluid internal
energy ${\mathcal E}=hu_t$ at the equator).

To model inhomogeneities in the interstellar medium,
we inject perturbations of varying amplitude to the (tangential-) velocity components at the outer boundary. 
Perturbations are modeled as time-varying Gaussian random field with a white noise spectrum

\begin{equation}
	\label{eq:injected_spectrum}
	S_{|\delta {\bf u}|} (k) \sim {\rm constant} \,,
\end{equation}
in the wavelength range $\lambda_k/M=2\pi/kM \in [214,2400]$,
and in the frequency range
$f_k \in [3.7,41] \times 10^{-5} M^{-1}$
(see Appendix~\ref{sec:boundary} for more details).
The remaining fluid variables at the boundary are set consistently with the initial
condition, and are therefore constantly injecting matter
that should preserve the initial state in absence of perturbations.
The injected noise is controlled by the parameter $\delta$ which specifies the ratio of the variance of velocity perturbations to the radial component of the 4-velocity of the unperturbed flow at the 
boundary, $\delta=\langle \delta {\bf u}^2 \rangle^{1/2}/u^r$.
We adopt the adiabatic index $\hat{\gamma}=4/3$.

We perform several simulations varying $\ell_0$ and $\delta$.
In order to isolate the effect of perturbations, we run
a set of simulations in a Bondi-Michel accretion scenario,
that is, $\ell_0=0$ and $a=0$.
The list of runs and parameters used is displayed in Table~\ref{tab:runs}.

To run the simulations, we use the code \texttt{BHAC} \citep{Porth2017,Olivares2019}.
We use a spherical polar grid in modified Kerr-Schild coordinates
with logarithmic spacing in radius.
The base resolution is
$N_r\times N_\theta \times N_\phi = 96\times 48 \times 48$ and
we employ 3 levels of \ac{AMR}, obtaining an effective resolution of $384\times 192 \times 192$.
The inner boundary is located inside of the central black hole event horizon, at $r=1.19\ M$,
in order to avoid boundary effects.
We employ a finite volume method with piecewise parabolic reconstruction
(PPM), a total variation diminishing Lax-Friedrichs (TVDLF)
approximate Riemann solver and a two-step method for time integration
\citep[see][for more details on coordinates and numerical methods]{Porth2017}.

To reduce the cost of simulations, we evolve a passive tracer $f$
that is initialized as $f=0$ inside the domain and $f=1$
for the injected matter at the boundary,
and evolve only those blocks of $8\times8\times8$ cells
for which $f>0.1$ or which are surrounded by blocks that satisfy this condition.
For all of the simulations, this tracer reaches the event horizon at
$t\lesssim 30\ 000\ M$, after which the simulation domain becomes active everywhere.
We continue the evolution up to
$t=60\ 000\ M$, which corresponds to a total simulation time of nearly
5 free-fall timescales from the sonic radius,
$t_{\rm ff}=\pi (r_s/2)^{3/2}\approx12\ 418\ M$.

\begin{table}
	\centering
	\label{tab:runs}
	\caption{Table of runs.}
	\begin{tabular}{l l l l}
	 \hline
	 $\delta $	& $\ell_0=0$ & $\ell_0=2.25$ & $\ell_0=3.25$ \\
	            & $a=0$      & $a=0.95$      & $a=0.95$      \\
	 \hline\hline
	 0.01         & \texttt{l0p001} & \texttt{l2p001} & \texttt{l3p001} \\
	 \hline
 	 0.1          & \texttt{l0p01}  & \texttt{l2p01}  & \texttt{l3p01}  \\
 	 \hline
 	 1            & \texttt{l0p1}   & \texttt{l2p1}   & \texttt{l3p1}   \\
 	 \hline
 	 10           & \texttt{l0p10}  & \texttt{l2p10}  & \texttt{l3p10}  \\
 	 \hline
	\end{tabular}
\end{table}

\subsection{Physical parameters}
\label{sec:phys-params}

Accretion onto an object at rest with respect to a spherically symmetric,
asymptotically uniform medium can be considered to start at the
Bondi radius, $r_{\rm B}$, the distance at which the asymptotic sound speed
$c_\infty$ equals the escape velocity, that is, $r_{\rm B}=2GM/c^2_\infty $.
Temperatures inferred from Chandra X-ray observations of the hot
gas surrounding Sgr A* \citep{baganoff_chandra_2003} and
the central black hole of M87 \citep{russell_inside_2015},
combined with the assumption of a monoatomic ideal gas with $\hat{\gamma}=5/3$,
yield estimates for the Bondi radius of $6\times 10^5\ M$ and
$4\times 10^5\ M$, respectively.
The several orders of magnitude separation between the Bondi radius and the
event horizon makes simulations of accretion from the Bondi radius onto
\acp{SMBH} prohibitive for most numerical codes.
In practice, due to temperature gradients, the {\it local} sound speed
$c_s$ does not coincide with escape velocity at the Bondi radius.
This happens instead at the sonic radius $r_s=2GM/c^2_s $,
which marks the transition from subsonic to supersonic flow.
In Newtonian hydrodynamics, the case $\hat{\gamma}=5/3$ is degenerate
and pushes the sonic radius to the origin.
However, by incorporating relativistic corrections and assuming $c_\infty \ll c$,
it takes a finite value that can be approximated
as $r_s \approx 3M c/4 c_\infty$ \citep[see e.g.][]{rezzolla_relativistic_2013}.
For the value of $c_\infty$ reported above, this corresponds to
$r_s \approx 409 M$ for Sgr A* and $r_s \approx 335 M$ for M87.
The value $r_s = 500 M$ in our simulations is chosen accordingly within
the same order of magnitude.


Following the same relativistic Bondi models, the dimensionless temperature at
the sonic radius can be estimated to be
$\Theta\coloneqq k_{\rm B}T / m c^2 = 7.3\times 10^{-4}$ -- $8.8\times 10^{-4}$,
where $m$ is the ion mass and $k_{\rm B}$ is the Boltzmann constant.
For monoatomic hydrogen, this corresponds to $T \approx 8\times 10^9$K --
$10^{10}$K (higher values correspond to M87).
The dimensionless temperatures attained at the sonic radius for our simulations
are similarly $\Theta \approx 8\times 10^{-4}$. Expecting it to
to increase by orders of magnitude when approaching the black hole,
we set $\hat{\gamma}=4/3$.
The effective adiabatic index in this regime is very dependent on uncertain factors
such as cooling and the ratio between ion and electron temperatures, and a more self-consistent
generation the background solution may require the use of a relativistic equation
of state as in \citet{Aguayo-ortiz2021}. However, a fully accurate modeling
of these effects is beyond the scope of this project.


Turning to the second parameter, $\ell_0$, the specific angular momentum from stellar stellar wind accretion can be roughly estimated by $\ell\simeq r_{\rm acc}^2\Omega/4$ \citep{FrankKingEtAl2002}. Here $\Omega$ is the orbital angular velocity of the star and $r_{\rm acc}=2GM/v_{\rm w}^2$ is the accretion radius for an assumed cold wind with velocity $v_{\rm w}$.  Scaled to geometric units and for a star in Keplerian orbit with semi-major axis $a$ we have
\begin{align}
    \ell \simeq 0.5 \left(\frac{a}{\rm pc}\right)^{-3/2}\left(\frac{v_{\rm w}}{1000\rm km\, s^{-1}}\right)^{-4} \, .
\end{align}
Thus low angular momentum flows are indeed expected for these fiducial values.  
Focusing on a particular source, it was argued in \citet{Moscibrodzka2006} that the stellar complex known as IRS 13 E3 \citep{MaillardPaumardEtAl2004} exerts the the strongest ram-pressure at the Galactic center which renders it the dominant wind accretion source.  
Thus taking IRS 13 E3 with fiducial wind velocity of $1000\rm km\, s^{-1}$ as exemplary case and using the orbital fits by \cite{MuzicSchodelEtAl2008},
we obtain $\ell$ in the range $0.1-16$. 
This large spread is caused by the large range of admitted semi-major axes $0.1\rm pc - 2.6\rm pc$ reported in \cite{MuzicSchodelEtAl2008}.  

As there are large uncertainties associated with the value of $\ell$ in the Galactic center and to gain insight into the parameter dependence, we here investigate three cases that correspond to the different qualitative behaviors of the
background solution: nonrotating case ($\ell=0$), a rotating case where the solution is complete,
that is, it connects smoothly infinity and the event horizon ($\ell=2.25$),
and a rotating case with an incomplete solution ($\ell=3.25$). Incomplete solutions of flows coming from infinity are expected either to pass through a shock and transition to another smooth solution that reaches horizon, or to represent flows that are unstable in absence of viscosity. For a sufficiently viscous flow, some of these incomplete solutions can transition to a torus \citep{Chakrabarti1996}. It should be noted that
a complete solution can exist for $\ell$ even when there is a circularization radius
$r_{\rm circ} > r_{\rm H}$ at which $\ell$ is equal to the Keplerian angular momentum, as it is the case for $\ell=2.25$ ($r_{\rm circ}\approx 3.6\ M$).
The reason is that fluid elements have a nonzero radial velocity and depending on their energy (part of which is internal) their centrifugal barrier is located further inside $r_{\rm circ}$, and in some cases they can even cross smoothly the event horizon.

Although \cite{ressler_ab_2020} showed that the orientation of the
flow angular momentum at horizon scales can vary wildly, these variations occur on a scale of hundreds of years
for Sgr A*. The simulations presented here have a much shorter duration -- comparable to 14 days for the same source -- which justifies the assumption that the orientation of the angular momentum is fixed.  

Finally, the most uncertain parameter is the amplitude of injected perturbations.
To explore several possibilities, we have considered a wide range with cases varying by
orders of magnitude with respect to the inflow velocity at the boundary.

\section{Results}
\label{sec:results}
\subsection{Global properties}
\label{sec:global-properties}

\begin{figure*}
	\includegraphics[width=\linewidth]{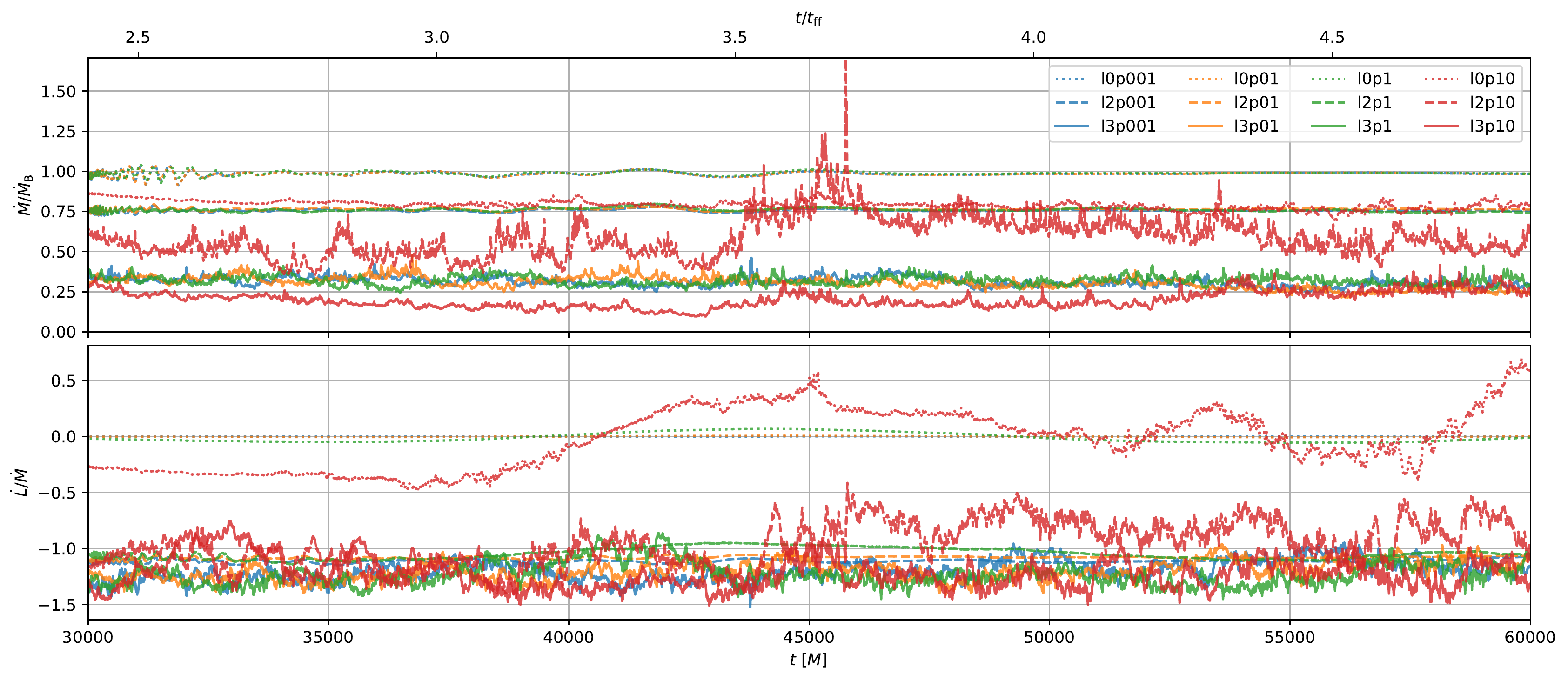}
	\caption{Mass and angular momentum flux through the event horizon for
	     all of the simulations, starting from a time where perturbations
	     have reached the event horizon for all simulations.
	     The upper horizontal scale measures time in units of the free-fall
	     timescale from the sonic radius.}
	\label{fig:horizon_fluxes_all}
\end{figure*}

In this section we briefly discuss and compare the salient global
features of the simulations.
We start by computing time series of the mass and angular momentum
flux through the event horizon

\begin{align}
\dot{M}(t)&\coloneqq \int_0^{2\pi} \int_0^{\pi} \rho u^r \sqrt{-g}\ d\theta\ d\phi \,, \\
\dot{L}(t)&\coloneqq \int_0^{2\pi} \int_0^{\pi} T^r_{\ \phi} \sqrt{-g}\ d\theta\ d\phi \,.
\end{align}
To set a typical scale that can be compared with real systems,
we normalize the accretion rate to the Bondi rate

\begin{equation}
	\dot{M}_{\rm B} = 4\pi \lambda_{\rm B} (GM)^2 \frac{\rho_\infty}{c^3_{\infty}}
\end{equation}
where
\begin{equation}
	\lambda_{\rm B} = \frac{1}{4}\left(\frac{2}{5-3\hat{\gamma}}\right)
	^{\frac{5-3\hat{\gamma}}{2(\hat{\gamma}-1)}} \,,
\end{equation}
and $\rho_\infty$ and $c_{\infty}$ are the density and sound speed at infinity.
In the units employed here, $\rho$ is normalized so that
$\rho=1$ at $r=6\ M$, which leads to the numeric value
$\dot{M}_{\rm B}\approx 246$ code mass units per gravitational timescale.

Figure \ref{fig:horizon_fluxes_all} shows the time series in the interval $t/M\in [30\ 000,60\ 000]$.
The first important feature shown in Figure \ref{fig:horizon_fluxes_all} is the long-term stability of the horizon penetrating fluxes over the simulated timescales.  While a quasi-stationary state is expected due the constant mass supply at the inflow boundaries, it is reassuring that accretion rates are nearly constant after $\sim 2$ freefall timescales.

As it could be expected, there is a trend that relates a higher angular momentum with a lower
mass accretion rate. A larger amplitude of perturbations also appears to result in
smaller accretion rates, likely due to the extra angular momentum provided by
perturbations, which also contribute to centrifugal support.
For instance, the addition of $\delta=10$ perturbations for the $\ell=0$ case
reduces the accretion rate to a value of $\sim 0.75\ \dot{M}_{\rm B}$ comparable to that obtained
for the simulations for $\ell=2.25$ with smaller perturbations.
The simulation with largest perturbation and angular momentum possesses the smallest accretion rate, at $\lesssim 0.25\ \dot{M}_{\rm B}$.


Inspecting the accretion of angular momentum, the solutions show a surprising behavior:
although it could be expected that a flow with larger angular momentum would result in
more angular momentum accreted by the black hole, the simulations with $\ell=2.25$
actually register slightly more angular momentum accretion than those with $\ell=3.25$.
Normalizing the angular momentum accretion rate by the mass accretion rate,
as it appears in the bottom panel of Figure \ref{fig:horizon_fluxes_all},
both cases show about the same value of $\dot{L}/\dot{M}$.
The reason is likely that centrifugal support prevents matter from accreting and
carrying angular momentum through the event horizon.
In this respect, it is important to recall the qualitative difference between the
unperturbed flow configurations corresponding to these two values: while $\ell=2.25$
allows solutions that connect smoothly infinity with the event horizon,
$\ell=3.25$ produces an incomplete solution which for the viscous case
should connect to a rotation supported torus where the flow is stalled
\citep{Chakrabarti1996}.

The time series in Figure \ref{fig:horizon_fluxes_all}
shows different variability properties for each simulation, with
those having higher $\ell$ and larger perturbations appearing more `noisy'.
For the cases with $\ell=3.25$, this can be attributed again to the fact that
the unperturbed solution is incomplete, producing shocks and complex interactions
between the flow close to the centrifugal barrier even when the injected perturbations are small.
However, it is interesting to see that the most variable time series corresponds
to $\ell=2.25$, for the simulation \texttt{l2p10}, where peaks
in $\dot{M}$ are sometimes even larger than the Bondi accretion rate.
We will diagnose the variability properties of the different solutions in more detail in Section \ref{sec:observable}.

Simulations with $\delta \leq 1$ and $\ell \leq 2.25$ show transient
oscillations near the time at which the innermost grids become active
($t\approx 30\ 000\ M$), and decrease in amplitude and frequency
as the evolution proceeds.
These are especially noticeable for the cases $\ell = 0$ for $\dot{M}$ and
$\ell = 2.25$ for $\dot{L}$. For the other cases, the
oscillations are masked by the larger perturbations.

\begin{figure*}
	\includegraphics[width=\linewidth]{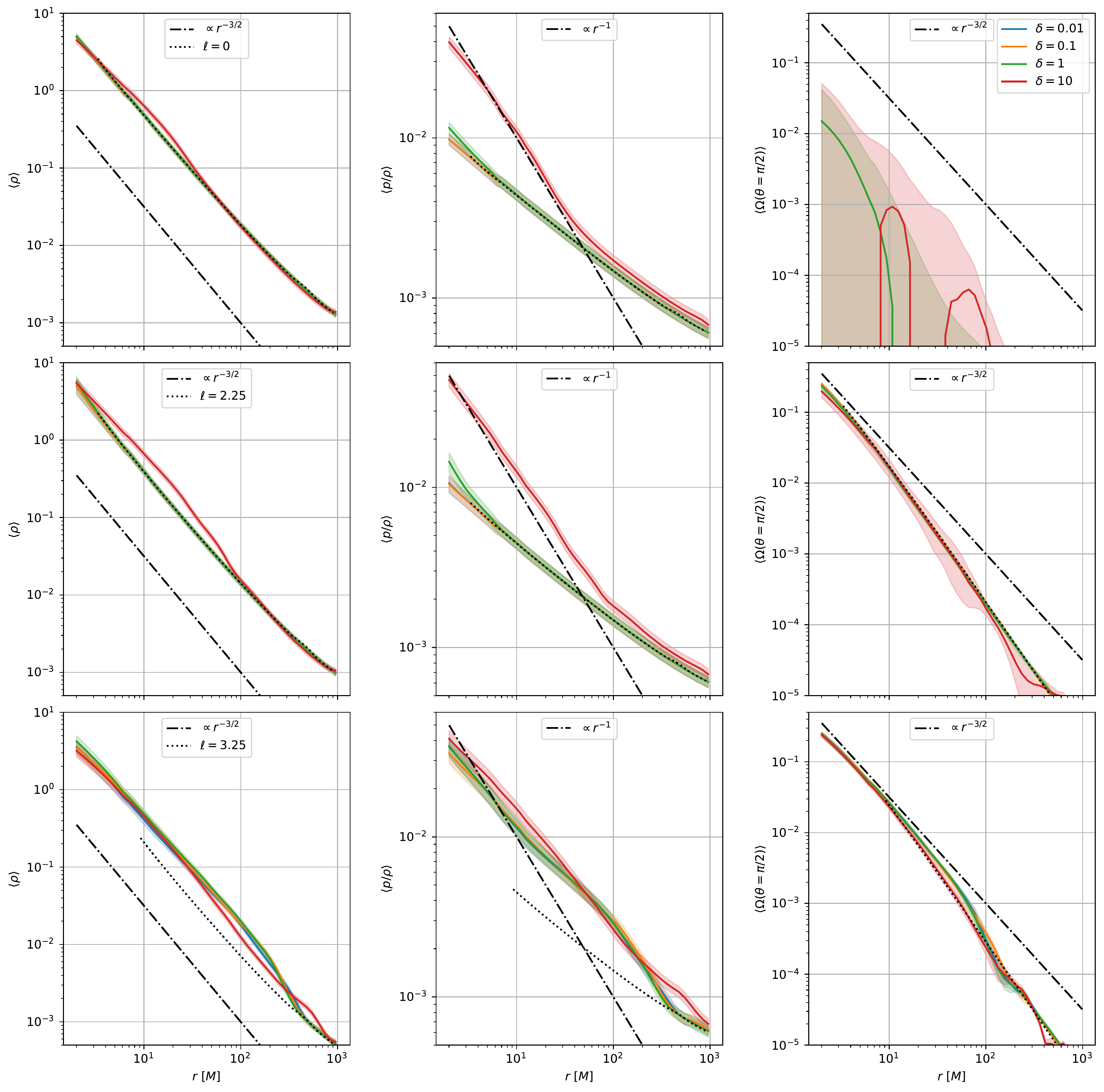}
	\caption{Radial profiles of density ({\it left column}),
		dimensionless temperature ({\it middle column}),
		and equatorial angular velocity ({\it right column})
		averaged over angle and time in the interval
		$t/M\in [50\ 000,60\ 000]$ for all simulations.
		From top to bottom, the columns correspond to
		$\ell_0=0$, $\ell_0=2.25$ and $\ell_0=3.25$, respectively.
		The shaded regions indicate the standard deviation.
		The dot-dashed lines show the power laws expected for an ADAF
		with $\gamma=4/3$ and no radiative cooling, and the dashed
		lines are the profiles for the unperturbed configurations with
		constant angular momentum used as initial condition.
		In most of the panels, the profiles for $\delta=0.01$ and
		$\delta=0.1$ overlap completely.
	}
	\label{fig:radial_profiles_all}
\end{figure*}

To quantify the departure of the perturbed solutions from the initial background solution, in Figure \ref{fig:radial_profiles_all} we show the angle- and time-averaged radial profiles for all simulations.

These are computed as

\begin{equation}
\label{eq:averaged-profiles}
\langle q \rangle (r,t) \coloneqq \frac{{\int_{0}^{2 \pi} \int_{0}^{\pi}} q (r,\theta,\phi,t) \ \sqrt{-g} \ d\theta \ d\phi}{\int_{0}^{2 \pi} \int_{0}^{\pi} \ \sqrt{-g} \ d\theta \ d\phi} \,,
\end{equation}
where $q=\rho,p/\rho$. We also show $\phi$-averages of the rotation
angular velocity on the equatorial plane $\Omega=u^\phi/u^t$,

\begin{equation}
\label{eq:omega-profile}
\langle \Omega \rangle (r,t) \coloneqq \frac{1}{2\pi}\int_{0}^{2 \pi} \Omega (r,\theta=\pi/2,\phi,t) \ d\theta \ d\phi \,,
\end{equation}
where axial symmetry has been used to eliminate the metric determinant.
All quantities are then time-averaged
over the interval $t/M\in [50\ 000,60\ 000]$.
We also plot the radial profiles expected for a self-similar ADAF
model \cite{Narayan1994} with $\hat{\gamma}=4/3$ and no radiative cooling,
as well as those of the unperturbed Chakrabarti solutions
with $\ell=0,\ 2.25$ and $3.25$ which are used as initial conditions
and are exact on the equatorial plane.

For all simulations, the density profiles (left column of Figure \ref{fig:radial_profiles_all})
are well described an ADAF profile of $\rho\propto r^{-3/2}$ which also holds approximately for the initial condition. In particular, the profiles are inconsistent with the shallower convective solution $\rho^{-1/2}$ indicating that convection is not important in our parameter regime \citep{NarayanIgumenshchevEtAl2000}.

Although the averaged density profiles shown in Figure \ref{fig:radial_profiles_all}
are smooth, some of the simulations exhibit density jumps at individual snapshots,
in which case the profile possess the same slope at either side
of the jump. These are present in the simulations
with high angular momentum and large perturbations.
As it will be discussed in Section \ref{sec:3D-morphology},
they are related to shocks which propagate outwards as
they are smoothed away.
Traces of these jumps are visible in the profiles of simulations
with $\ell=3.25$ (especially of {\texttt{l3p10}})
at scales of $r=10^2$ -- $10^3\ M$.
These are expanding shocks produced by the fluid colliding with
the centrifugal barrier, as those studied for example by \cite{Sukova2017}.
The slow evolution timescales near the outer boundary prevent them from
being smoothed by the time average.

The averaged profiles of $p/\rho \propto \Theta$ in the central
column of Figure \ref{fig:radial_profiles_all} show a more interesting behavior.
While initially they coincide with those of the Chakrabarti solutions, those
corresponding to $\ell=3.25$ and $\delta = 10$ gradually transition to the
profile of the ADAF solution. The clearest case is that of simulations
\texttt{l0p10} and \texttt{l2p10},
which transition form the constant-entropy initial profile
($\propto r^{-1/2}$ for the Bondi solution)
to that of the ADAF model $\propto r^{-1}$ at $r\approx 40\ M$.

The rise of the temperature profile close to the black hole
is likely a result of heating by shocks and turbulence, which transform
to thermal energy the kinetic energy injected through the perturbations
in the velocity.
It can be noticed that the temperature profiles of
\texttt{l0p1} and \texttt{l2p1} start rising as well and
deviate from the initial profile at a shorter distance from the black hole.
This suggests that indeed the radius at which the transition to an ADAF-like
profile occurs is related to the amplitude of perturbations in the medium.
The fact that \texttt{l0p10} acquires an ADAF-like temperature profile close to
the black hole is interesting. In fact, this simulation differs from the scenario
studied by \cite{Narayan1994} from which the self-similar solution is derived.
Here there is no coherent disk-like structure and the average of $\Omega$
is close to zero (see rightmost panel of Figure \ref{fig:radial_profiles_all}).
It is therefore surprising that the heating provided by incoherent shocks
and turbulence results in a temperature profile similar to that of a coherent
viscous rotating flow.

The rightmost column of Figure \ref{fig:radial_profiles_all} shows the angular
velocity profile for all simulations.
As expected, the profiles
corresponding to the cases with zero angular momentum in the unperturbed solution show negligible rotation velocity on the
equatorial plane.
The rest of profiles behave in a similar way as those of dimensionless temperature:
they follow the Chakrabarti constant angular momentum profile at large radii (yielding a power law slope of $-2$ far from the black hole)
and transition to the ADAF-like Keplerian profile $\propto r^{-3/2}$ closer
to the black hole.

In general, it appears that at large distances the system is well described by the
adiabatic Bondi- and Chakrabarti-like solutions, while once perturbations
become enough amplified by the geometry of the flow to produce shocks and turbulence,
entropy production starts and the system becomes better described by
ADAF-like profiles.

\subsection{Three-dimensional morphology}
\label{sec:3D-morphology}

\begin{figure*}
	\includegraphics[width=0.9\linewidth]{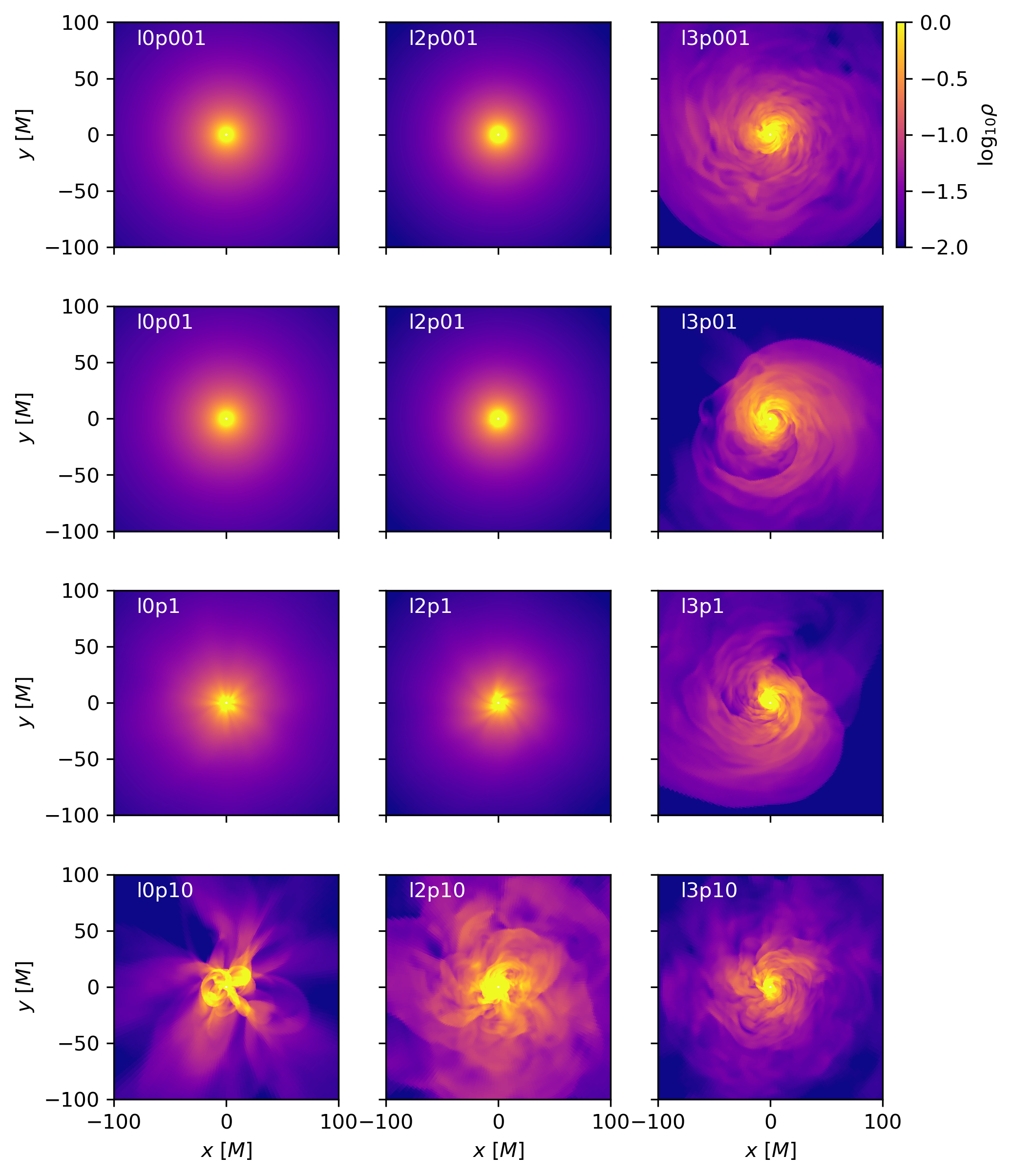}
	\caption{Logarithmic density maps for all simulations at $t=60\ 000\ M$
	on the equatorial plane. Panels are organized in the same way as simulations in Table~\ref{tab:runs}, that is,
	increasing angular momentum from left to right, and amplitude of perturbations from
	top to bottom.
	Movies of simulations \texttt{l0p10}, \texttt{l2p10}, and \texttt{l3p10},
	are available at 
	\url{https://youtu.be/1TQV_aX13xE},
	\url{https://youtu.be/oOh2reL9yK0}, and
	\url{https://youtu.be/VmCc3ZnDxEM}, respectively. }
	\label{fig:rho_equatorial}
\end{figure*}

\begin{figure*}
	\includegraphics[width=0.9\linewidth]{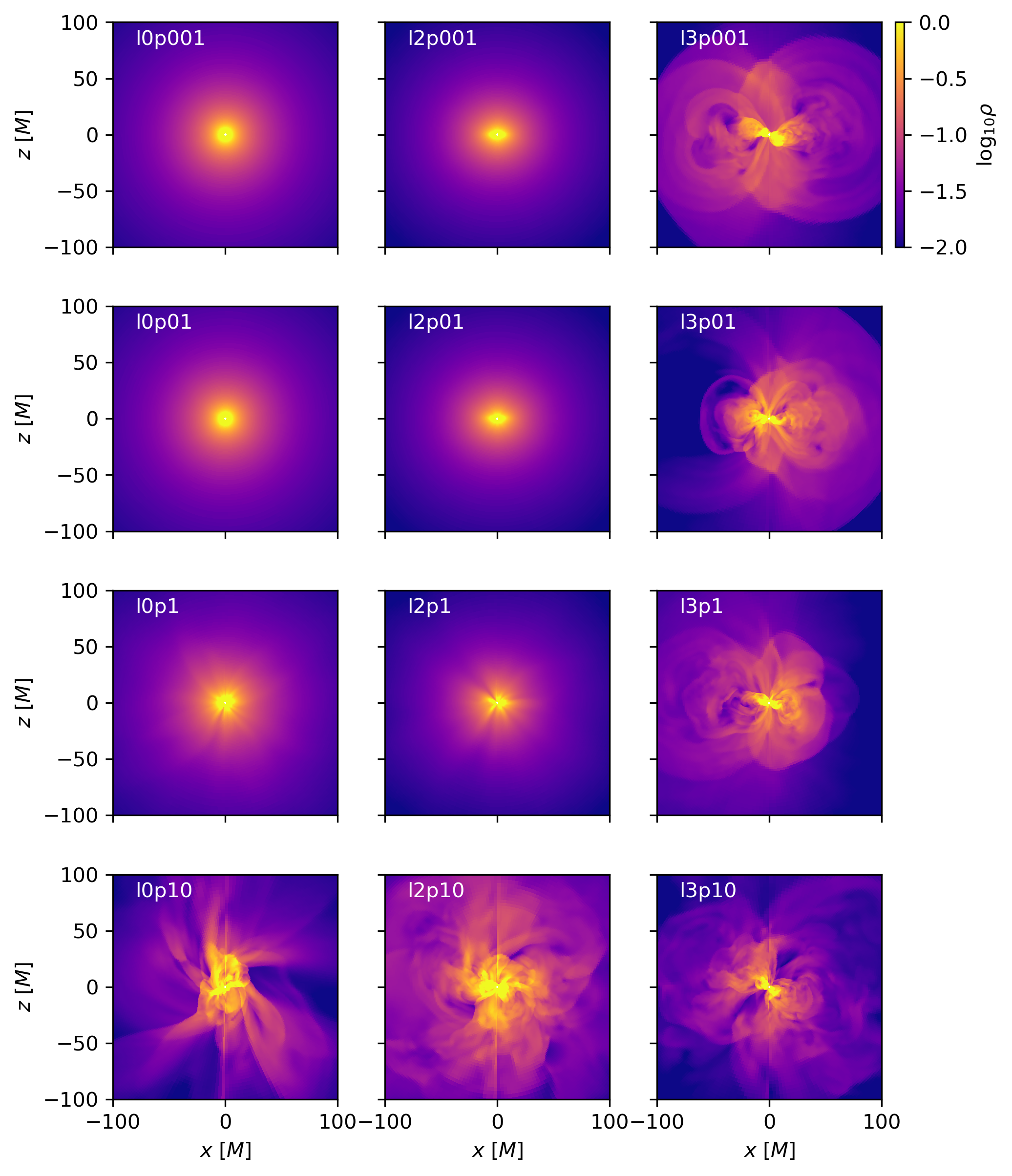}
	\caption{Similar as Figure \ref{fig:rho_equatorial}, for the meridional plane.}
	\label{fig:rho_meridional}
\end{figure*}

Figures \ref{fig:rho_equatorial} and \ref{fig:rho_meridional}
show the density distribution for all the simulations on the equatorial and the meridional
plane, respectively, at $t=60\ 000\ M$.
The four simulations with lowest $\ell$ and perturbation amplitude, \texttt{l0p001},
\texttt{l2p001}, \texttt{l0p01}, \texttt{l2p01} are always smooth and highly symmetric,
and are practically indistinguishable from the initial conditions.
Changes start becoming visible for those with perturbations comparable to the
incoming radial velocity, \texttt{l0p1} and \texttt{l2p1}, for which near-radial filaments can be seen.

The simulations with higher angular momentum are qualitatively different,
as it could be expected due to the incompleteness of the Chakrabarti solution.
In all of the $\ell=3.25$ runs it is possible to see the formation of a turbulent
torus-like structure
close to the black hole. The larger the perturbations are, the more misaligned this
structure becomes with respect to the large-scale angular momentum, which points into
the $+z$ direction (rightmost panels of Figure \ref{fig:rho_meridional}).

Despite the apparent similarity of these configurations with those in BHT simulations,
they exhibit important differences. While for BHT simulations the toroidal structure
is confined mainly by the equilibrium between gravity and the centrifugal force,
for the simulations presented here it consists in large part of unbound outflowing matter
that is confined by its interaction with inflowing matter.
In addition, while for BHT simulations most of the accretion flow occurs in the equatorial plane,
here the toroidal structure is an obstacle for the inflowing matter, causing most of the
accretion to occur through the poles.
This behavior is consistent with that observed for similar systems in absence
of magnetic fields, for example, by \citet{Proga2003a,Moscibrodzka2008,Sukova2017}.
We expect as well that the inclusion of magnetic fields will reverse the
situation by producing accretion on the equatorial plane and a polar outflow \citep{Proga2003b}.

\begin{figure*}
	\includegraphics[width=0.9\linewidth]{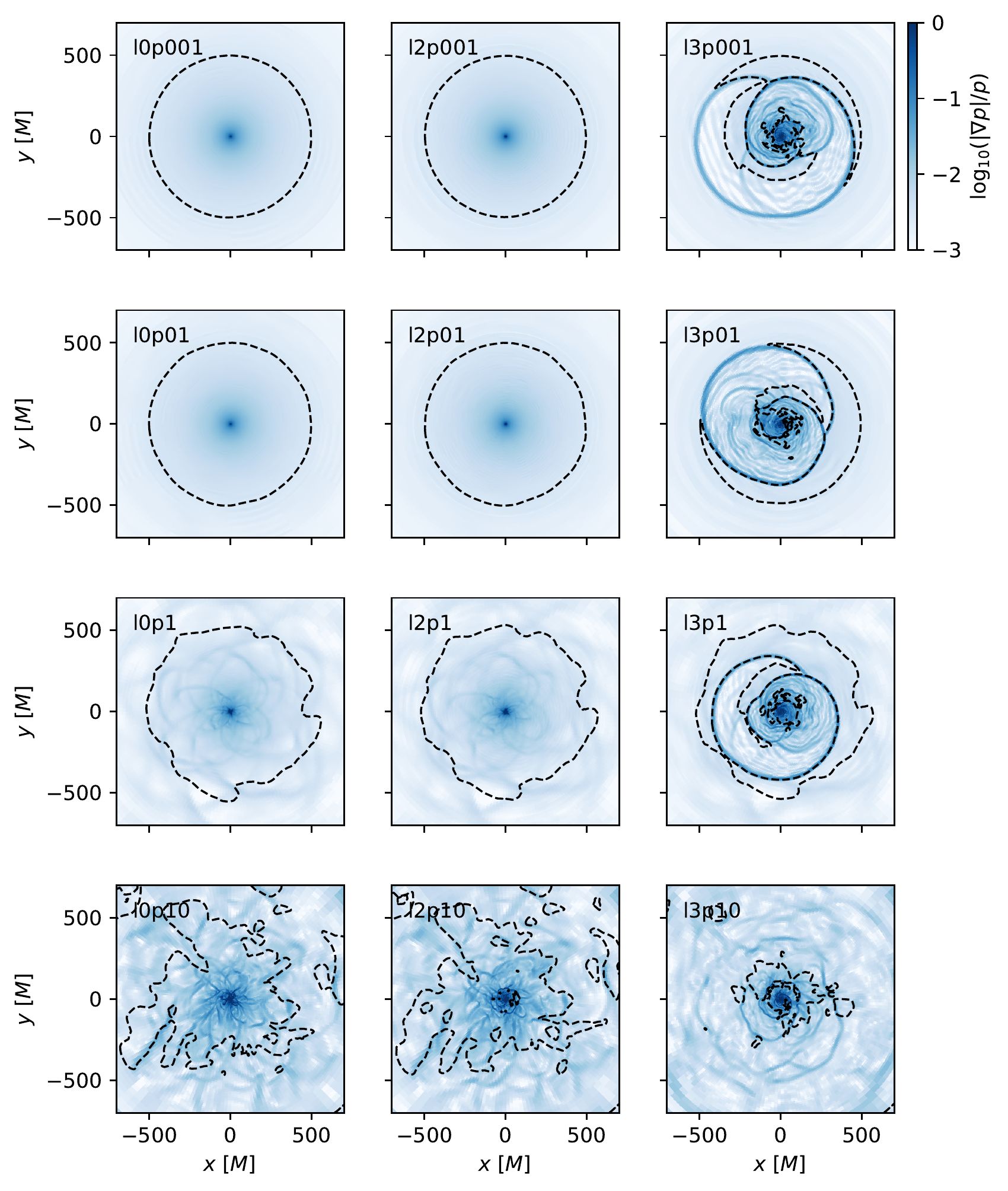}
	\caption{Shocks and sonic surfaces for all simulations at $t=60\ 000\ M$.
	    The color scale displays the relative pressure gradient,
	    which is used as a proxy for shock locations,
		and the dashed lines indicate the static limits of the
		sonic metric.}
	\label{fig:gradp_equatorial}
\end{figure*}

\begin{figure*}
	\includegraphics[width=0.9\linewidth]{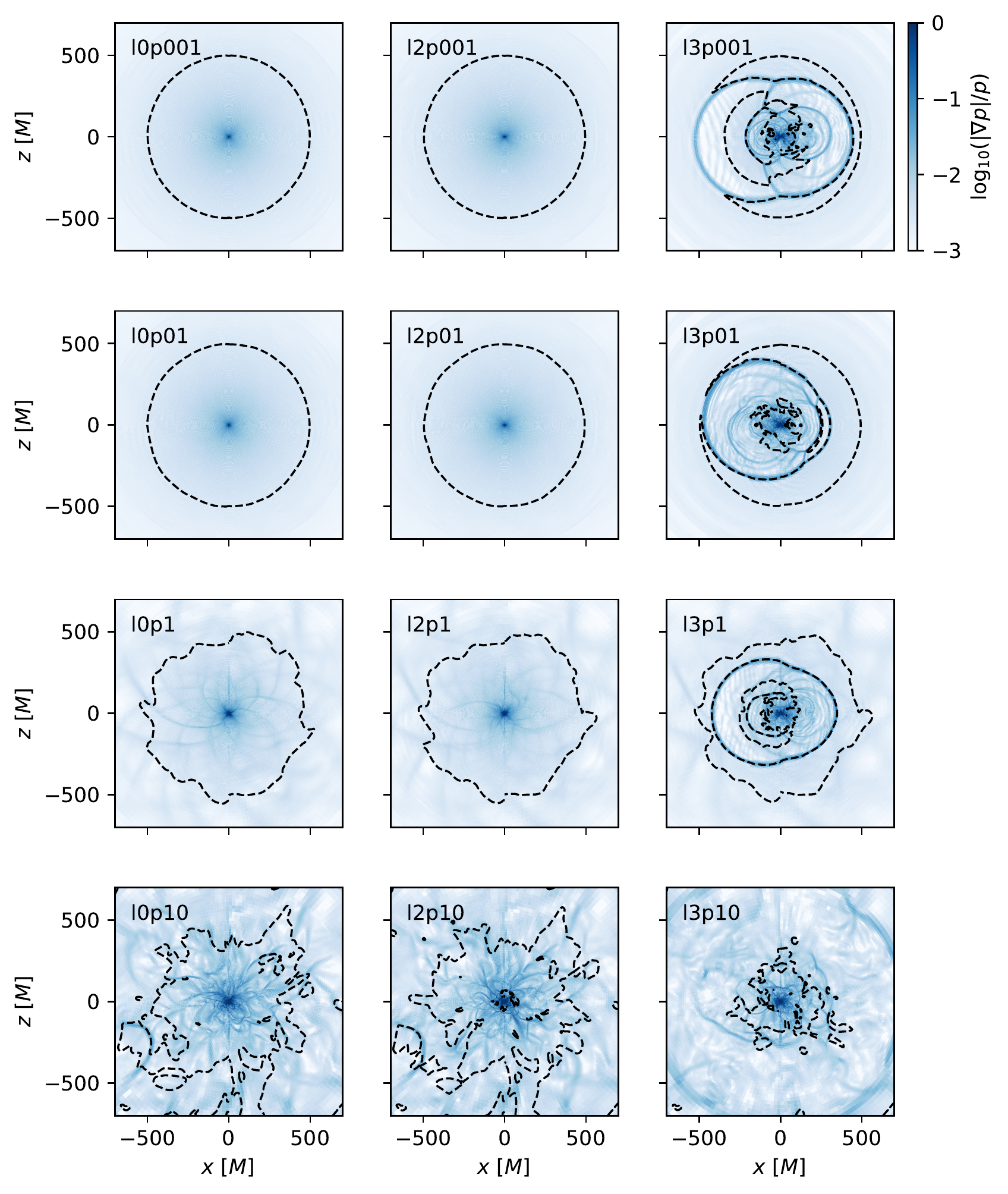}
	\caption{Similar as Figure \ref{fig:gradp_equatorial},
	for the meridional plane.}
	\label{fig:gradp_meridional}
\end{figure*}

Figures~\ref{fig:gradp_equatorial}
and \ref{fig:gradp_meridional} show the relative pressure gradient
as proxy for the location of shocks.
The simulations with low $\ell_0$ and $\delta$, (\texttt{l0p001},
\texttt{l2p001}, \texttt{l0p01}, \texttt{l2p01}) do not show important pressure gradients.
In contrast, those with $\ell_0=3.25$ and $\delta\leq 1$ (\texttt{l3p001}, \texttt{l3p01},
and \texttt{l3p1}) show clear spiral shocks 
This coherent large scale shock does not form in the
strongly perturbed case $\delta=10$.
The colormap in Figures \ref{fig:gradp_equatorial}
and \ref{fig:gradp_meridional} also allow to see sound waves traveling within the shocked regions.

It is also interesting to examine to what extent the causal structure of the flow
is preserved in presence of perturbations and high angular momentum.
The dashed lines in Figures~\ref{fig:gradp_equatorial} and
\ref{fig:gradp_meridional} mark the surfaces for which
the 4-velocity of an observer at rest at infinity,
$\partial_t=(1,0,0,0)$, becomes null with respect to the sonic metric
\citep{Moncrief1980},

\begin{equation}
	\mathcal{G}_{\mu\nu}= \frac{\rho}{h c_s}\left[g_{\mu\nu} + (1 - c_s^2) u_\mu u_\nu\right] \,,
\end{equation}
where $\rho$ is the rest-mass density, $h$ is the enthalpy, $c_s$ is the
sound speed and $u_\mu$ is the four velocity of the fluid.
This condition is analogous to that defining static surfaces such as ergoregions
and event horizons for the spacetime metric $g_{\mu\nu}$, and can be used
to characterize transitions between subsonic and supersonic flows in an invariant
way \citep{Aguayo-ortiz2021}, especially in situations that lack symmetries such
as the perturbed flows studied here.

The sonic surface for the unperturbed solutions is a sphere with radius $r_s=500\ M$,
centered at the black hole.
Its structure is practically unchanged for the four cases with lowest $\ell_0$
and $\delta$, (\texttt{l0p001}, \texttt{l2p001}, \texttt{l0p01}, \texttt{l2p01}).
Cases with $\ell_0\leq 2.25$ and $\delta=1$ (\texttt{l0p1}, \texttt{l2p1})
exhibit slight but noticeable changes in the shape of the sonic surface,
although it remains close to $r_s$.
This is remarkable since perturbations already have an amplitude
similar to the magnitude of the inflow radial velocity at the boundary.
Only when the perturbation amplitude is ten times the inflow radial velocity
(\texttt{l0p10}, \texttt{l2p10}) we see large incursions of subsonic matter inside
$r_s$, as well as islands of supersonic (subsonic) flow within the former subsonic
(supersonic) regions.

Models with $\ell_0=3.25$ show a more complex causal structure.
The spiral shock produces an additional transition from supersonic to subsonic
flow, and it can erase the original sonic surface as it propagates outwards.
However, downstream the flow can become supersonic again.
The spiral structure can then produce several sonic transitions between the
distant regions, where matter is injected subsonically, and the event horizon,
that needs to be crossed supersonically.
For instance, in the panel of Figure~\ref{fig:gradp_equatorial} that corresponds
to simulation \texttt{l3p1}, there can be even five sonic transitions
when approaching the black hole from certain directions.
For the case with $\delta=10$, \texttt{l3p10}, the original sonic surface
has disappeared completely, and a new one has formed closer to the black hole.
Also in this case, the spiral shock produces more than one sonic transition
in some directions.

\begin{figure}
	\includegraphics[width=1.0\linewidth]{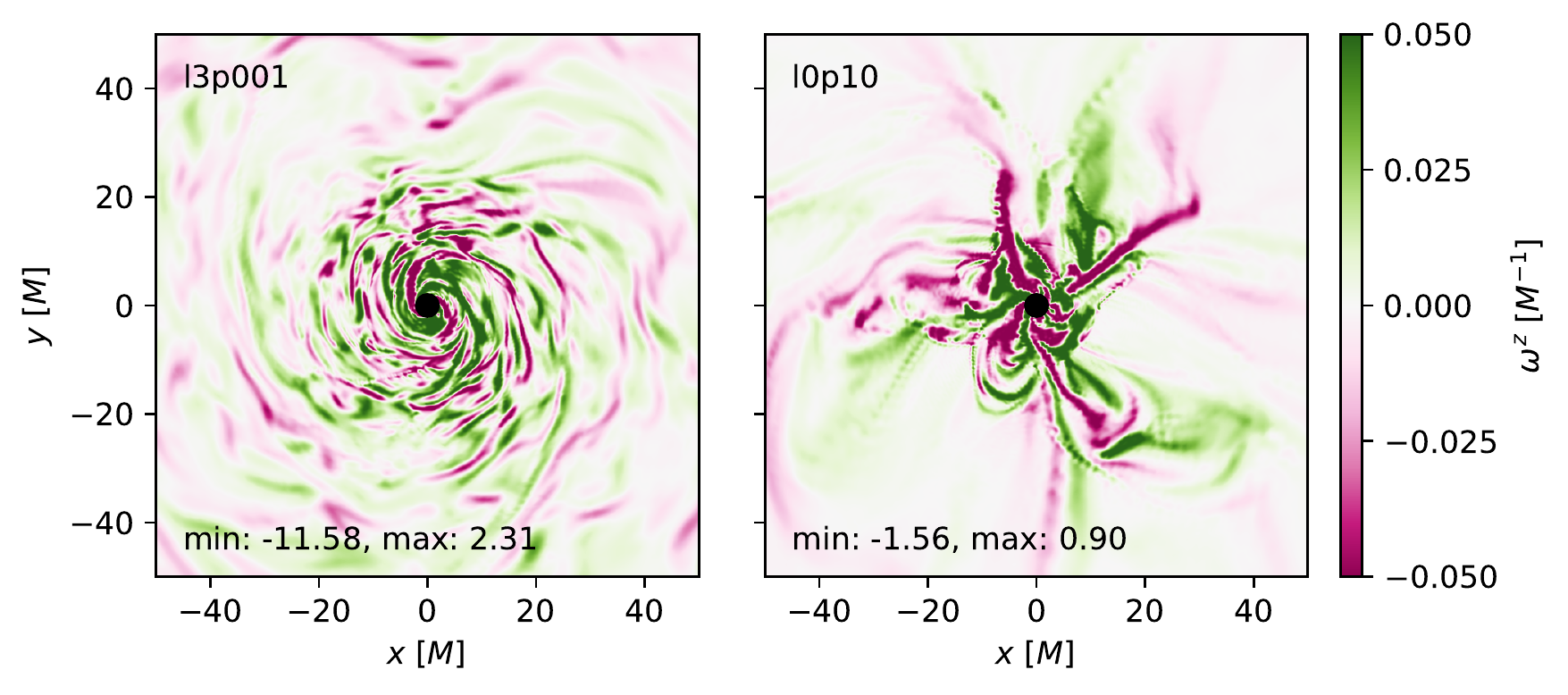}
	\caption{Vertical component of the vorticity
	on the equatorial plane for simulations \texttt{l3p001}
	and \texttt{l0p10}.}
	\label{fig:vorticity}
\end{figure}

In addition to the entropy increase due to shocks, turbulence could also play
a role in heating the fluid and contribute to the transition from a constant entropy temperature profile to an ADAF-like profile, as shown in the middle panel of Figure~\ref{fig:radial_profiles_all}.
In order to highlight the presence of turbulence, Figure~\ref{fig:vorticity}
shows the $z$-component of the vorticity vector on the equatorial plane for
simulations \texttt{l3p001} and \texttt{l0p10}.
The $\ell_0=0$ case shows vorticity sheets that can be associated with
fluid streams approaching the black hole at different speeds, and suggest
the emergence of smaller turbulent structures if simulated at a higher resolution.

For the incomplete $\ell_0=3.25$ analytical solution
the fluid is expected to form a torus at the circularization radius
$r_{\rm circ} \approx 8.8 M$, and no presence of fluid is expected
at smaller radii. However,
in all of our simulations we observe the flow occupying this region without impediment,
meaning that a means of angular momentum redistribution is operating.
The same can also be inferred from the rotation velocity profiles in the rightmost
panel of Figure \ref{fig:radial_profiles_all}, where several of them transition
from constant to Keplerian angular momentum profiles. This indicates that even in the absence
of magnetic fields, and thus MRI and large scale Maxwell stresses,
angular momentum redistribution occurs, and it can be attributed to shocks
and turbulence that could easily appear in nature.


\subsection{Variability properties}
\label{sec:observable}

To have a rough estimation of the observable properties of the variability
in our simulations, we have computed synthetic X-ray light curves by
integrating the total bremsstrahlung emissivity from free-free
electron-ion collisions \cite{rybicki_radiative_1986},

\begin{equation}
	\epsilon_{\rm BR} = 5.54 \times 10^{-9}\
	Z^2\ 
	\left(\frac{m_{\rm i}}{m_{\rm p}}\right)^{1/2}
	\left(\frac{n_{\rm i}\ \rho}{10^{6}\ {\rm cm}^{-3}}\right)^2
	\left(\frac{p}{\rho}\right)^{1/2}\ \frac{{\rm erg}}{{\rm cm}^3\ {\rm s}}\,,
\end{equation}
where $Z$ is the atomic number of ions,
$m_{\rm i}$ is the ion mass, and $m_{\rm p}$ is the proton mass, and
$n_{\rm i}$ is scaling factor that relates the dimensionless
code density $\rho$ with the ion number density $n_{\rm i} \rho$.
The Gaunt factor has been assumed to be constant and equal to $1.2$.
The integration is performed as

\begin{equation}
L_{\rm BR} = 2.41 \times 10^{38}\
                   \left(\frac{M}{4.15\times 10^{6}\ M_{\odot}}\right)^3
                   \int \epsilon_{\rm BR}\  \Gamma \sqrt{\gamma}\; {\rm d}^3 x\ \frac{\rm erg}{\rm s}\,,
\end{equation}
where $\Gamma$ is the Lorentz factor, and the prefactor comes
from the conversion of volume in geometrized code units to physical units.

\begin{figure*}
	\includegraphics[width=\linewidth]{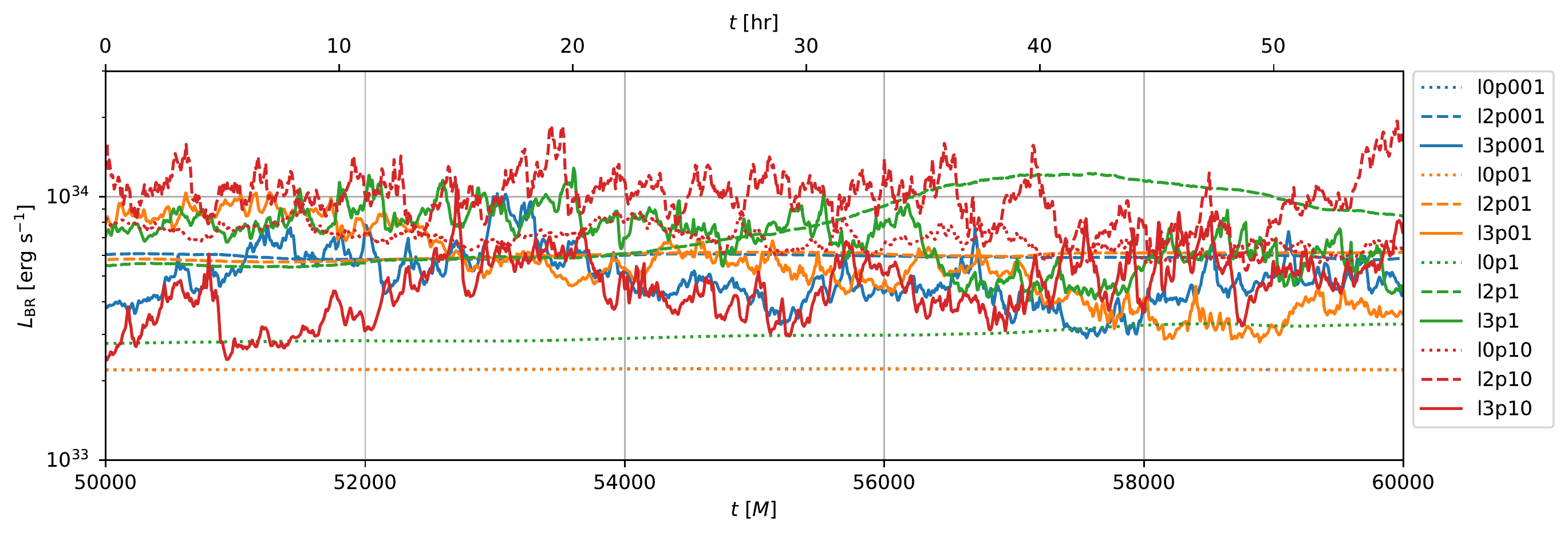}
	\caption{Synthetic light curves of total Bremsstrahlung luminosity using the parameters of Sgr A*,
		for all simulations. The curve corresponding to \texttt{l0p001} overlaps completely to that of
	    \texttt{l0p01}. The upper horizontal axis displays the time in hours.}
	\label{fig:L_brems_all}
\end{figure*}

\begin{figure}
	\includegraphics[width=\linewidth]{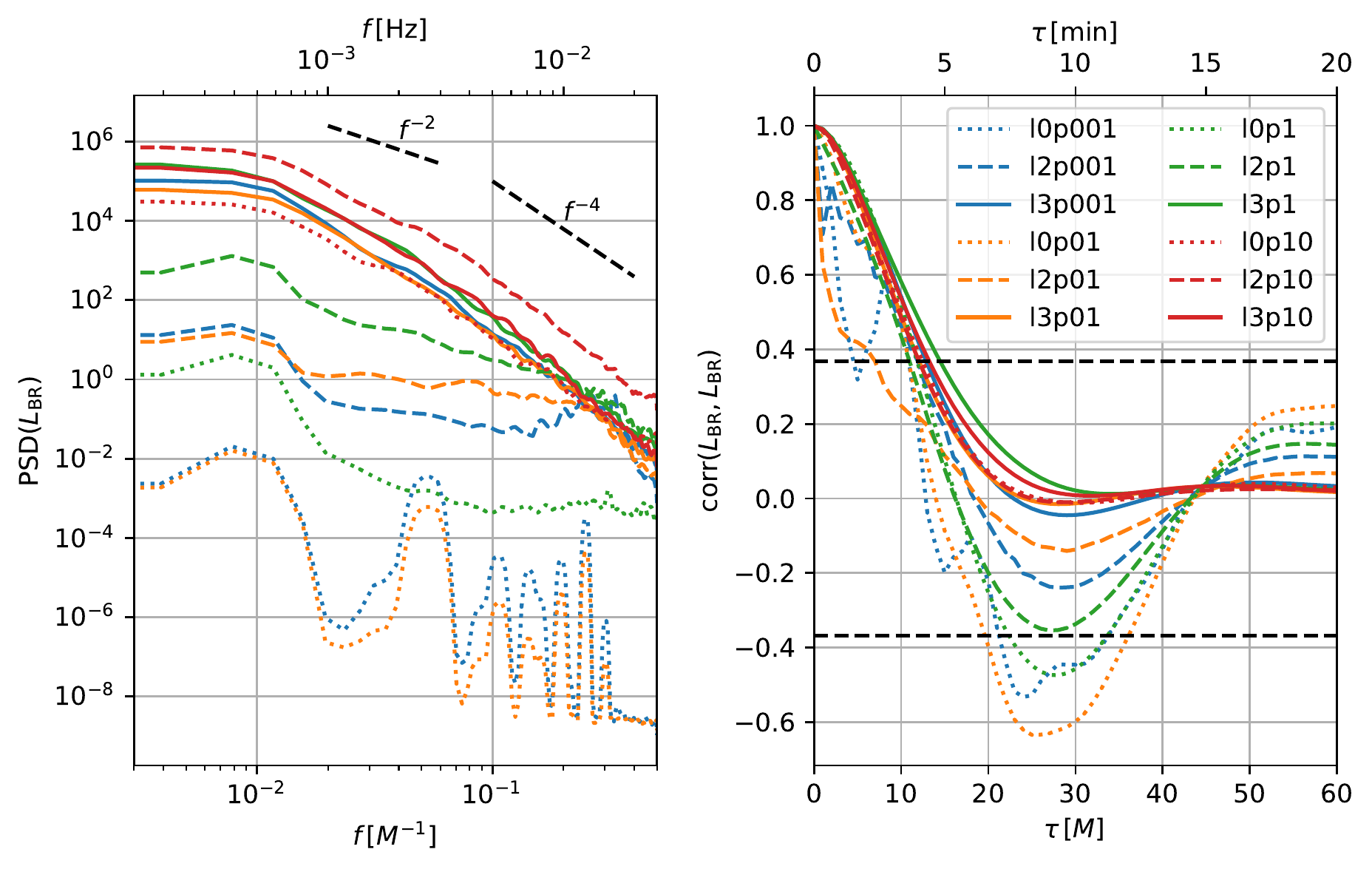}
	\caption{Spectrograms ({\it left}) and correlation functions ({\it right})
		of the bremsstrahlung light curve proxy of all simulations.
	    Dashed lines with slopes of power laws are shown for comparison
	    in the left panel, and bound the region between $\pm 1/e$ in the
	    right panel. The upper horizontal axes have been scaled
        for Sgr A*.}
	\label{fig:spectrogram_correlations}
\end{figure}

The synthetic light curves within $t/M\in[50\ 000, 60\ 000]$ for each simulation are shown
in Figure \ref{fig:L_brems_all}.
The parameters have been chosen for monoatomic hydrogen,
with $n_{\rm i} = 10^6\ {\rm cm}^{-3}$, and $M$ as the mass of Sgr A*,
$M=4.15\times 10^6 M_{\odot}$.
This gives luminosities that agree in order of magnitude with the
$\approx 2.4\times 10^{33} {\rm erg\ s^{-1}}$ estimated by \cite{baganoff_chandra_2003}.
For this source, the time interval corresponds to $\approx 55.5$
hours of observing time. We calculated spectrograms in this interval
using the Welch method \citep{welch_use_1967}
with time windows overlapping over 128 points ($=128M$).
The \acp{PSD} are shown in the left panel of Figure \ref{fig:spectrogram_correlations}.
It can be seen that, as expected, the power of fluctuations increases with the
amplitude of the injected perturbations and with the angular momentum.
The two simulations with zero angular momentum and smallest perturbations
show small frequency peaks that are lost into the noise for the other cases.

As perturbations and angular momentum increase, it is possible to observe
a steepening in the slope of the \ac{PSD}.
Simulations with $\ell_0=3.25$ or $\delta =10$ show a very similar
spectrum with a break from white noise to red noise around $f \sim 10^{-2} M^{-1}$.
Power laws of red noise $f^{-2}$ and $f^{-4}$ are shown for comparison.
Spectrograms are calculated over a frequency range higher than that of
the injected perturbations (see Secion \ref{sec:setup}),
which therefore do not appear in the \ac{PSD}.

We Fourier-transformed these spectrograms in order to obtain autocorrelation
functions ${\rm corr} (L_{\rm BR},L_{\rm BR})$ for the synthetic lightcurve.
For all of the simulations, positive correlations decay below $1/e$
in about $\tau \sim 5$ -- $15\ M$. However, autocorrelations for noisier simulations
($\ell_0=3.25$ or $\delta=10$),
are close to zero for $\tau \sim 30\ M$ ($\approx 10$ minutes for Sgr A*),
while simulations with small angular momentum and perturbations still exhibit
longer term positive and negative correlations.

Overall, the correlation timescales are shorter than 40 $M$ for all simulations,
which allow us to calculate modulation indices $\sigma/\mu$ for statistically
uncorrelated data by computing the standard deviation $\sigma$ and average $\mu$
over points separated by 50 $M$. Modulation indices are shown in Table \ref{tab:modulation_idx}.
The modulation index of the mass accretion rate through the event horizon in the same time
interval is shown in parenthesis.
The fact that the latter doesn't show as much variations when the former varies by
orders of magnitude indicates that an important portion of Bremsstrahlung variability
is not related to fluctuations in the accretion rate close to the horizon.
Instead the Bremsstrahlung modulation index clearly increases for those simulations
in which shocks and turbulence are present.

\begin{table}
	\label{tab:modulation_idx}
	\caption{Modulation index of the bremsstrahlung luminosity light curve
		and the mass accretion rate through the event horizon (in parenthesis),
		computed over the interval $t/M=[50\ 000,60\ 000]$. }
	\begin{tabular}{c c c c }
		$\delta u_{\rm RMS}/|u^r| = $	& $\ell=0$ & $\ell=2.25$ & $\ell=3.25$ \\
		\hline\hline
		0.01        & 0.004 (0.003) & 0.016 (0.002) & 0.246 (0.066) \\
		\hline
		0.1         & 0.003 (0.003) & 0.028 (0.002) & 0.319 (0.101) \\
		\hline
		1           & 0.057 (0.001) & 0.303 (0.012) & 0.237 (0.075) \\
		\hline
		10          & 0.105 (0.031) & 0.225 (0.031) & 0.269 (0.200) \\
		\hline
	\end{tabular}
\end{table}

\begin{figure*}
	\includegraphics[width=0.9\linewidth]{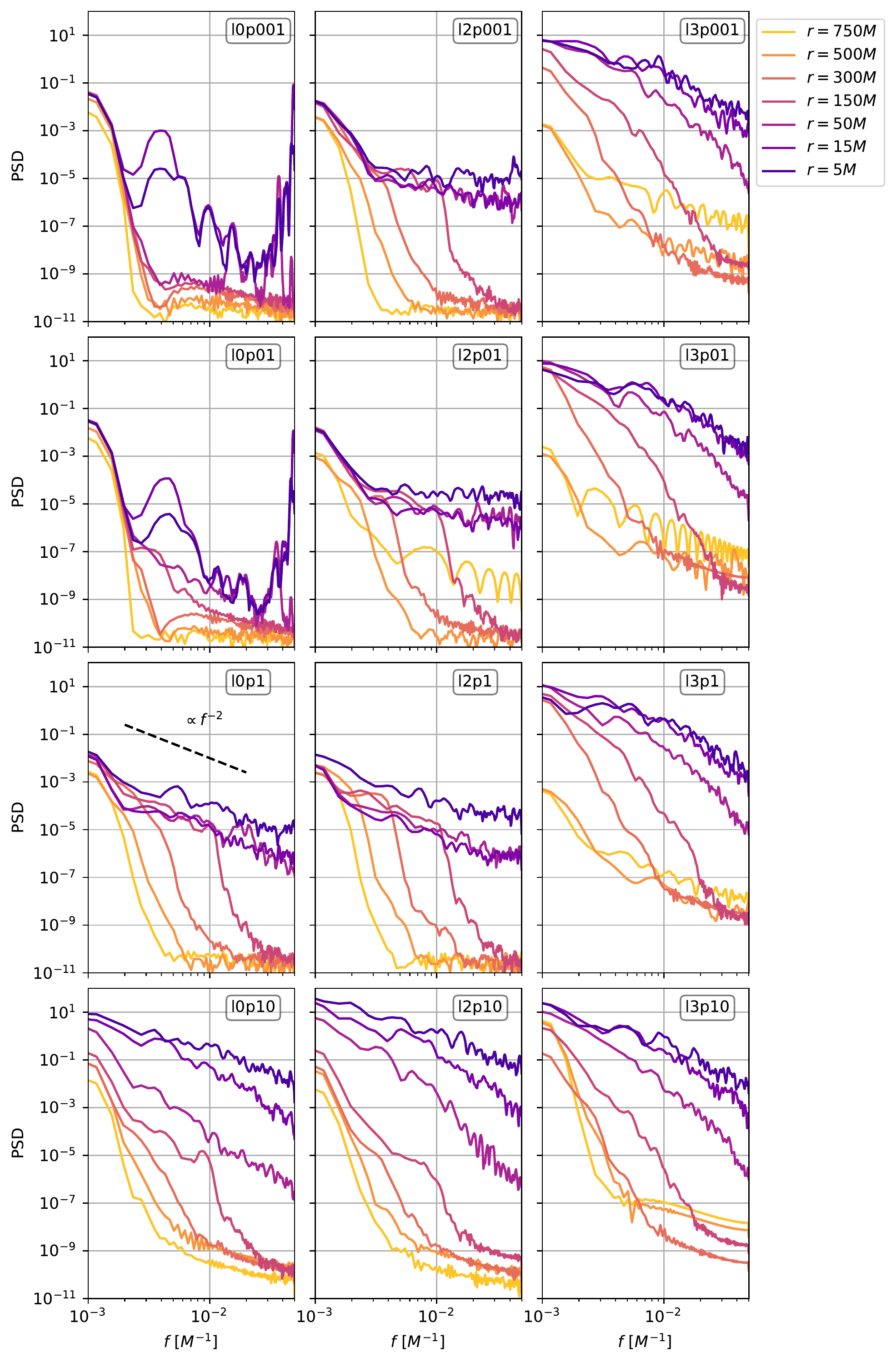}
	\caption{Power spectral density of the mass accretion rate
		measured at different radii for all of the simulations,
		during the interval $t/M\in[50\ 000, 60\ 000]$.
	    The dashed line represents a power law $\propto f^{-2}$.}
	\label{fig:spectrograms_radial}
\end{figure*}

In order to investigate the origin and properties of the fluctuations
observed in the mass accretion rate, we calculated \acp{PSD}
of $\dot{M}(r,t)$ at several radial shells.
These are shown for selected radii and for all the simulations in
Figure~\ref{fig:spectrograms_radial}, for the same interval used
for the analysis of the synthetic Bremsstrahlung light curve
an using the same methodology as in Section~\ref{sec:observable}.

It is evident that in general the spectrum at event-horizon
scales differs significantly from that at large distances.
In most of the panels it is possible to see higher frequencies
increasingly populated as one moves closer to the black hole.
This can be interpreted as the transfer of energy from longer lower
frequency modes to smaller and faster modes, and could be due to
the change in the characteristic scale of the system as the
fluid moves inwards, as well as to the development of turbulence.
Cases producing shocks ($\ell_0=3.25$ and $\delta>1$) show
a larger power, and the spectrum at the smaller radius acquires the form
of a power law $\propto r^{-2}$ or steeper. The rest of cases with $\ell_0=2.25$
show a flatter spectrum close to white noise.
Similarly as for Bremsstrahlung spectrograms,
for cases with $\ell_0=0$ and $\delta=0.01,\ 0.1$ the spectrum of $\dot{M}$
has a very small power and is dominated by oscillation peaks.
These appear within $r\leq 15\ M$, indicating that, for almost unperturbed
Bondi-like accretion, fluctuations in Bremsstrahlung emission
do appear related to these fluctuations in  $\dot{M}$ close to the horizon.

In contrast, our results suggest that for sufficiently large perturbations,
as those injected manually or as those produced by the movement
of the spiral shock, a red noise spectrum will be recovered,
regardless of the injected perturbation spectrum.

\section{Discussion and Conclusions}
\label{sec:discussion}

In this work we used 3D \ac{GRHD} simulations to study the structure and
variability patterns in perturbed transonic accretion flows with low-angular momentum.
Our aim was to explore an accretion scenario that generalizes torus
simulations, is more consistent with the properties of stellar wind-fed accretion,
and is controlled by a reduced set of parameters.
Our simulation setup also aims to overcome two of the known limitations
of the \ac{BHT} simulation paradigm, namely, the secular decrease in torus
mass that complicates long-term variability studies and the artificiality
of the medium beyond the close vicinity of the black hole.
We evaluated the general properties of this accretion scenario using
several diagnostics, namely,
(1) time series of the mass accretion rate
and angular momentum flux through the event horizon,
(2) shell- and time-averaged profiles of several quantities of interest,
and (3) a synthetic Bremsstrahlung light curve used to analyze
its variability properties.
We also investigated the 3-dimensional morphology of the models,
including the location of shocks and sonic surfaces.

Our models, contrary to \ac{BHT} simulations, have accretion rates that do not
decay exponentially, allowing for long-term variability studies.
We observe that $\dot{M}$ decreases significantly for models with larger angular
momentum and perturbation amplitude.
This is consistent with the additional centrifugal support provided by angular
velocities and the additional pressure support due to heating caused by turbulence
and shocks, that are also more important for the same models.
The reduction in mass accretion rate for larger angular momentum models
also results in a smaller net angular momentum flux through the horizon,
suggesting that there is a finite value of $\ell_0$ that maximizes
the accretion of angular momentum.

The fact that these models are fed from solutions that extend to
infinity allows to relate the mass accretion rate at horizon scales to the
Bondi accretion rate from the medium properties at large scales.
This in turn permits to obtain tighter constraints on the models.
For example, density scales need to match large scale fluid properties
in addition to electromagnetic flux constraints derived from radiative
transfer calculations.

In addition to their significant variations in accretion rate,
the flows we studied possess a rich phenomenology and are in some respects qualitatively
different both from \ac{BHT} simulations and from unperturbed transonic flows.
Some of the salient features we observe include outflowing toroidal structures,
turbulence, shocks, filaments, and multiple sonic transitions.


Deviations from the transonic solutions used as initial conditions
are in some cases large enough to lead
to different averaged temperature and velocity profiles.
In particular, models with large perturbations and angular momentum
deviate from the isentropic temperature profiles and transition to
profiles similar to those of an \ac{ADAF}.
For the cases initialized from complete solutions ($\ell_0=0$ and $\ell_0=2.25$),
the radius at which the transition occurs seems related to the amplitude of perturbations.
This can be explained from the instability of supersonic spherical accretion
to non-radial perturbations \cite{kovalenko_instability_1998}.
These perturbations grow without limit with smaller radii,
producing the `ADAF transition' when they become sufficiently
large to produce shocks and generate entropy, which depends on the injected
perturbation amplitude.

For the cases initialized from the incomplete solutions with small perturbation amplitudes,
large scale spiral shocks are produced by the interaction between the inflowing fluid and the centrifugal
barrier. They likely play an important role in
transporting angular momentum outwards thus enabling accretion of matter
and the transition to a Keplerian rotation profile at small radii in the absence of a magnetic field.

In the context of chaotic cold accretion \citep[e.g.][]{GaspariTemiEtAl2017,PrasadSharmaEtAl2017},
it has been suggested that inelastic collisions between clouds can cancel angular momentum,
leading to an increased accretion rate.
In the simulations presented here, shocks could be expected to play a similar role;
however, even simulations where shocks are present show the same trend that relates
larger perturbations and angular momentum with smaller accretion rates.
The reason could be that shocked simulations are precisely those with larger perturbations and
angular momentum, and this effect needs to compete with the additional support provided by
angular velocities and the pressure from gas heated by shocks and turbulence.


As could be expected, the different qualitative behavior
results in different variability properties of the simulations.  
However, we found that for sufficiently large angular momentum and perturbations,
a red noise spectrum is robustly recovered even from a white noise injection perturbations.


Our simulations are complementary to similar hydrodynamical simulations carried out by other authors. 
\citet{Ressler2018} uses a conservative hydrodynamics code to study the formation of the accretion
flow onto Sgr~A* in which matter is constantly supplied by stars on orbits around the central black hole.
This setup leads to a somewhat chaotic accretion. Although their simulation domain is much
larger in comparison to ours, their inner boundary overlaps with our outer boundary.
\citet{Ressler2018} obtained solutions with density and temperature power-law profiles
$\propto r^{-1}$ which is different compared to our results.
Also, the angular momentum in our simulations is significantly lower than the value they obtain
at comparable radii (In their Figure~14 \cite{Ressler2018} reports that the
$\ell \approx 0.4-0.5 \ell_{\rm K}$ at the inner boundary).
These differences in the profiles could be explained by the differences in the physical scenario
considered. In their case, these include stronger rotation, the presence of important outflows,
as well as line and Bremsstrahlung cooling, which becomes important at the scales they consider.
Similarly, very large scale simulations performed by \citet{guo_toward_2022} study the formation
of the accretion pattern at event horizon scales following material from the Bondi radius
scale in elliptical galaxies such as M87. The information obtained from these works and
future large scale simulations can be incorporated to smaller scale simulation setups as those
presented in this work, e.g. by specifying initial density profiles and the spatial and
temporal spectrum of injected perturbations.

In a setting more similar to ours, \citet{Sukova2017} uses 2D and 3D conservative \ac{GRHD}
simulations to study low angular momentum flows on closer to horizon scales,
but without manually injecting perturbations.
In the tests we performed while implementing our
initial condition we have recovered the general behavior of some of their 2D models,
although
there were some differences in implementation and parameter choices,
that we describe in Appendix~\ref{sec:background}.


The inclusion of magnetic fields in the simulations
will be presented in a forthcoming publication;
however, there are a few expectations we can draw from our
hydrodynamic models that could be relevant for observations.
For example, the fact that the accretion pattern
is almost isotropic for cases with low
angular momentum $\ell_0 \leq 2.25$ may results in images that are also
independent from orientation, in particular
contrast to \ac{SANE} \ac{BHT} models.
In addition, the possibility of producing synthetic
synchrotron lightcurves that do not suffer from
secular torus depletion may contribute to
some extent to unravel the ongoing discussion
on the suitability of \ac{GRMHD} models
to met the tight variability constraints
given by 230 GHz observations of Sgr~A*
\citep{CollaborationAkiyamaEtAl2022}.

Overall, we believe our simulations are an important middle-step towards
obtaining more realistic models of relativistic accretion flows
in which matter is supplied by the turbulent interstellar medium.

\section*{Acknowlegdements}
We thank Jesse Vos, Aristomenis Yfantis, Alejandra Jimenez-Rosales, Christiaan Brinkerink and other members of the EHT group at Radboud University for discussions. We also thank Jordy Davelaar and Agnieszka Janiuk for their comments.
HROS was supported part by a Virtual Institute of Accretion (VIA) postdoctoral fellowship from the Netherlands Research School for Astronomy (NOVA).
We acknowledge that the results of this research have been achieved using the DECI resource Snellius based in the Netherlands at SURF with support from the PRACE AISBL. This work made use of the following software libraries not cited in the text: MATPLOTLIB \citep{Hunter:2007} and NumPy \citep{harris2020array}. This research has made use of NASA’s Astrophysics Data System.

\bibliographystyle{aa}
\bibliography{references}

\appendix
\section{Initial conditions}
\label{sec:background}

Our initial data is constructed from the semi-analytic rotating
transonic solutions by \citet{Chakrabarti1996}.
These can be thought of as a generalization of Michel accretion \cite{Michel1972}
for a rotating flow and for the Kerr metric.

To solve for the fluid properties, one assumes that streamlines are radial when
projected on the meridional plane (that is, $\theta$-components of the velocity are neglected).
Mass flux $\sqrt{-g}\ \rho u^r$, entropy, internal energy $\mathcal{E}=h u_t$
and angular momentum $\mathcal{L}=-hu_\phi = \ell\mathcal{E}$
are conserved along the streamline.

Similarly as for Michel and Bondi accretion,
once $\mathcal{L}$ is specified, $\mathcal{E}$ can be chosen so that
the sonic radius is at the desired position.
The flow configuration is then found by solving a pair of coupled nonlinear algebraic equations for the sound speed and the radial velocity in the co-rotating frame
at every point \citep[equations 30a,b of][]{Chakrabarti1996}.
The equations do not constrain the density scale, which can be chosen later.
In our case, we set it so that $\rho=1$ at $r=6\ M$.

Accretion solutions of these equations have a wide variety of qualitative behaviors.
A class of solutions connects smoothly infinity and the event horizon
similarly as the Michel solution. Other solutions possess incomplete interior
or exterior branches that can be connected by a shock, and there exist also incomplete
solutions that have a sonic point but do not extend supersonically to the event horizon
\citep[see Figure 2 and Section 4.1 of ][for a complete description]{Chakrabarti1996}.
For simplicity, we always solve only the exterior branch of the solution.

To initialize our simulations, we solve the system on a grid
covering the range $\theta \in [0,\pi/2]$ with 300 points.
The supersonic part of each streamline was solved with 300 points and the subsonic
part with 100 points, both logarithmically spaced in radius.
The calculated part of the subsonic region extends beyond the
simulation domain by 10\% in order to be be used for the boundary conditions.

The approximation of projected radial streamlines is, in general, inconsistent with 
vertical equilibrium, however it holds on the equatorial plane.
For this reason, we choose an angular momentum profile with a sharp peak at the equator
that decays to $\ell=0$ at the poles, where again vertical equilibrium is fulfilled (equation \eqref{eq:ell-profile}).

\begin{figure*}
	\centering
	\includegraphics[width=\linewidth]{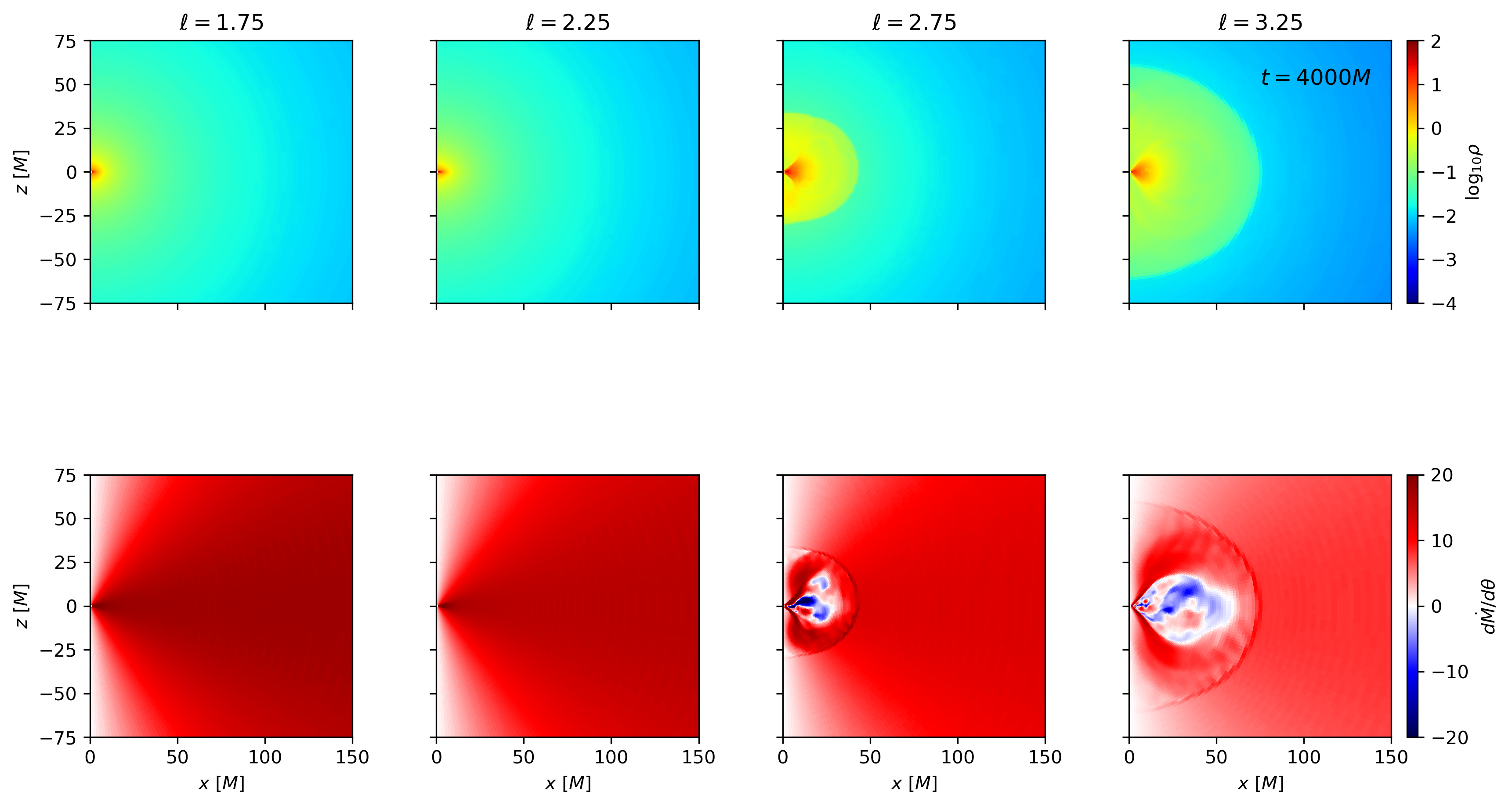}
	\caption{Rest-mass density and mass accretion rate per $\theta$ angle
		for 2D evolutions of unperturbed initial conditions with different angular momentum.}
	\label{fig:2D_bg_solutions}
\end{figure*}

\begin{figure}
	\centering
	\includegraphics[width=\linewidth]{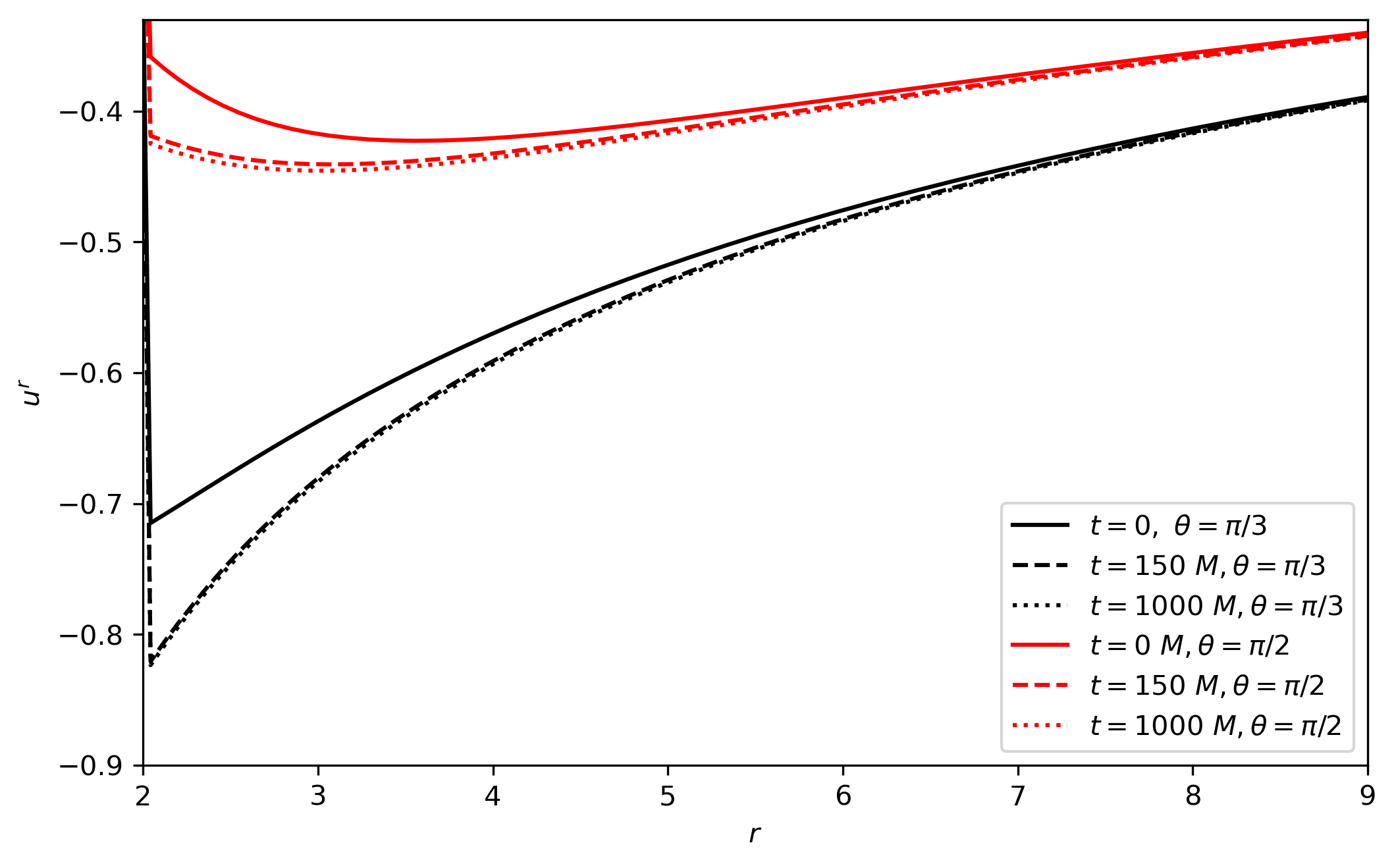}
	\caption{Equatorial cuts of the radial velocity of background flow
		at different times and latitudes for the case $\ell=2.25$.
		\label{fig:background_ur}}
\end{figure}

In order to evaluate the adequacy of this approximation and to test that the solution reproduced
the expected qualitative behavior, we performed 2D simulations
evolving only the initial condition without injecting any perturbation.
Figure \ref{fig:2D_bg_solutions} shows 2D maps of rest-mass density and
mass accretion rate per $\theta$-angle for some of the 2D simulations
we performed for $a=0.95$ and different angular momenta.
As expected, for those cases in which the equatorial solution
connects the event horizon and infinity ($\ell=1.75,2.25$),
the artificial initial condition quickly relaxes to a true solution
in vertical equilibrium and remains stable until the end of the simulation.
Figure \ref{fig:background_ur} shows the evolution of the radial 4-velocity profile
of the unperturbed solution used for simulations with $\ell=2.25$ at different latitudes.
At  $\sim 150\ M$, the configuration has already relaxed to a stationary flow.
This is sufficiently adequate for our simulations, which have durations more than 100 times longer.

For incomplete solutions ($\ell=2.75,3.25$),
material accreting on the equatorial plane starts piling up due to the centrifugal barrier,
while accretion continues through the poles. The dense toroidal structure that forms
is different from the tori commonly use in simulations in that it consists of
`outflowing' unbound material, which produces a shock when interacting with the
incoming accretion flow. Since there is no cooling,
these structures grow until the end of the simulation \citep{MolteniSponholzEtAl1996}.
As described in Section \ref{sec:3D-morphology}, these shocks are present as well in 3D simulations; however, the lack of azimuthal symmetry introduces important differences, such as the presence of turbulence in the $\phi$-direction and the change in shape of the shock from spheroidal to spiral.

Our 2D unperturbed simulations show a behavior that is consistent with that reported by
\citet{Sukova2017}. A difference with respect to that work is that their initial condition
is a Bondi flow to which rotation has been added, while in our case it is a solution
including rotation in a more self-consistent way (albeit exact only on the equatorial plane and the poles),
for which the assumption of low angular momentum is not necessary.
Another difference is that \citet{Sukova2017} is largely focused on studying the parameter
regime that produces oscillating shocks, which we have not explored.

\section{Boundary conditions}
\label{sec:boundary}

To emulate the turbulent flow entering from the boundary,
we filled the ghost zones with the same transonic solution used
as initial condition and added time-dependent noise in the form of \acp{GRF}.
These fields are generated as the superposition of plane waves with
random phases.

The usual way of generating a time-dependent \ac{GRF}
in $N$ dimensions is by Fourier-transforming white noise in $N+1$
dimensions, multiplying the Fourier transform by the desired \ac{PSD}
and then transform back. One can then `play' the time dependent noise
by successively applying slices of the resulting $N+1$-dimensional
array.

Although this way of generating \acp{GRF} is very fast due to the elegance of the
Fast-Fourier Transform (FFT) algorithm, it has some disadvantages
that lead us to follow a different procedure.
First, storing a large 4-dimensional
array and communicating it among different parallel processes to perform interpolations
can be a source of implementation and performance problems, and second,
we do not desire to apply the noise in a full three-dimensional box,
but only on the outer ghost cells, which make an almost two-dimensional spherical
shell embedded in three-dimensional space. 

For this reason, we build the \ac{GRF} by directly evaluating a series of
sine functions corresponding to plane waves with random phases at the cells of interest.
This sum has the form

\begin{equation}
	{\rm GRF}(t,x^i) = \sum_{k_1, k_2, k_3=-N_k}^{N_k} A_{\vec{k}}
	                   \sin\left[\frac{2\pi}{\lambda_{\rm max}}
	                             \left( \vec{k}\cdot\vec{x} - 	f(\vec{k})\ t 
	                             - \varphi_{\vec{k}} \right) \right]\,,
\end{equation}
where $\vec{k}$ is a three-dimensional vector of integers,
$\lambda_{\rm max}$ is the maximum wavelength,
$f(\vec{k})$ is a function determined by a user-defined dispersion relation,
and $\varphi_{\vec{k}}$ is the random phase corresponding to $\vec{k}$.
We draw random phases from a uniform distribution over $[0,1)$ using a pseudo-random
number generator with a fixed seed, which allows to use the same random phases
without the need to store them between restarts.
In order to ensure causality, we use the constant dispersion relation
$f(\vec{k})=c_s$, where $c_s$ is the sound speed at the simulation boundary,
although more complicated relations are also possible.
The coefficients $A_{\vec{k}}$ are set according to the desired
power spectral density $S_{\vec{k}}$, as $A_{\vec{k}} \propto (S_{\vec{k}})^{1/2}$, and can be normalized
to give the desired rms perturbation amplitude.
We set $A_{\vec{k}}=0$ for $\vec{k}=(0,0,0)$
in order to keep the average of the \ac{GRF} to zero.

In our simulation, we generate 3 \acp{GRF} to perturb the three spatial
components of the 4-velocity in Cartesian coordinates. We then transform
them to the code coordinates and add them only to the angular components
of the velocity given by the background transonic solution.

When giving up on the FFT, we pay the price of having to evaluate
a large number of transcendental functions, which can slow down the code significantly.
We find that $N_k = 5$ gives a negligible slow down and the smallest-wavelength
mode has a size comparable to $\sim 1.6$ cells of the outer boundary.

This leads to 
$N_\theta \times N_\phi \times N_{\rm ghost} \times N_{\rm fields} \times (2N_k+1)^3=12~266~496$ evaluations of the sine function per time step,
or $340~736$ evaluations for each of the $8\times8\times8$
\ac{AMR} blocks at the boundary. Here, $N_{\rm ghost}=4$ is the number
of ghost zones in the radial direction and $N_{\rm fields}=3$,
since each of the spatial components of the velocity is perturbed with a different \ac{GRF}.

Finally, it is worth mentioning that, being a sum of periodic functions,
the noise is also periodic with the period of the longest wavelength mode.
Although in general this will not result in a periodic behavior of the simulation
due to the changing chaotic dynamics inside the domain, it may be desirable to
produce non-periodic noise models. One possibility could be changing slightly the
dispersion relation so that the period of some of the modes is an irrational
multiple of that of others. The search for more appropriate non-periodic noise models
is, however, out of the scope of this work.


\end{document}